\documentclass[12pt]{iopart}

\usepackage{iopams}  

\makeatletter
\newcommand{\overset}[2]{\ensuremath{\mathop{\kern\z@\mbox{#2}}\limits^{\mbox{\scriptsize #1}}}}
\newcommand{\underset}[2]{\ensuremath{\mathop{\kern\z@\mbox{#2}}\limits_{\mbox{\scriptsize #1}}}}
\makeatother

\usepackage[colorlinks=true,linkcolor=blue,urlcolor=blue,citecolor=blue]{hyperref}
\usepackage{bbold}
\usepackage{tikz}
\usetikzlibrary{tikzmark}
\usepackage{braket}

\newcommand{\tikzsymbol}[2][circle]{\tikz[baseline=-0.5ex]\node[inner
sep=3.5pt,shape=#1,draw,#2]{};}%

\newcommand{\ws}{\tikzsymbol[rectangle]{minimum width=2pt,fill=white}}
\newcommand{\bss}{\tikzsymbol[rectangle]{minimum width=2pt,fill=black}}
\newcommand{\ess}{\tikzsymbol[rectangle]{minimum width=2pt,draw=white}}
\newcommand{\s}{\hspace{0.01in}}
\newcommand{\us}{\underset}

\makeatletter
\newcommand\xleftrightarrow[2][]{%
  \ext@arrow 9999{\longleftrightarrowfill@}{#1}{#2}}
\newcommand\longleftrightarrowfill@{%
  \arrowfill@\leftarrow\relbar\rightarrow}
\makeatother

\begin{document}

\title[Integrability breaking in the Rule 54 cellular automaton]{Integrability breaking in the Rule 54 cellular automaton}

\author{Javier Lopez-Piqueres$^1$, Sarang Gopalakrishnan$^2$ and Romain Vasseur$^1$}

\address{$^1$Department of Physics, University of Massachusetts, Amherst, Massachusetts 01003, USA \\
$^2$Department of Physics, The Pennsylvania State University, University Park, PA 16820,
USA}
\ead{jlopezpiquer@umass.edu}
\vspace{10pt}
%\begin{indented}
%\item[]November 2021
%\end{indented}

\begin{abstract}
Cellular automata have recently attracted a lot of attention as testbeds to explore the emergence of many-body quantum chaos and hydrodynamics. We consider the Rule 54 model, one of the simplest interacting integrable models featuring two species of quasiparticles (solitons), in the presence of an integrability-breaking perturbation that allows solitons to backscatter. We study the onset of thermalization and diffusive hydrodynamics in this model,  compute perturbatively the diffusion constant of tracer particles, and comment on its relation to transport coefficients.

\end{abstract}

%\maketitle

\section{Introduction}
\label{Sec: intro}
% TODO: write your article here.
Understanding the emergence of hydrodynamics in quantum and classical systems from simple, reversible, microscopic dynamics has been a long-standing enterprise~\cite{spohn2012large}. Cellular automata take a very special role in this direction as they represent an attempt to reduce to  \textit{Boolean logic} rules the equations of hydrodynamics \cite{rothman2004lattice}. Their use in this field has a long history going back to the 70s and extending through the 80s \cite{hardy1973time,PhysRevA.13.1949,PhysRevLett.56.1505,wolfram1986cellular,demasi1989hydrodynamics} with the culmination of a particular cellular automaton successfully modeling the Navier-Stokes equations governing viscous fluids~\cite{PhysRevLett.56.1505}. 

Cellular automata have made a come back in recent years, as toy models to explore questions regarding thermalization and ergodicity breaking in quantum systems~\cite{PhysRevLett.119.110603,gopalakrishnan2018facilitated,iaconis2019anomalous,iadecola2020nonergodic,feldmeier2020anomalous,iaconis2021multipole}, as well as for serving as minimal models of interacting (integrable or not) systems \cite{pozsgay2021yang,gombor2021integrable, gombor2021superintegrable, prosen2021reversible,prosen2021many}. Rule 54, a cellular automaton (CA) which involves two species of solitons which are distinguished by their chiralities (we will refer to them as `$R$' and `$L$' solitons, for \textit{right} and \textit{left} moving solitons) and which features many similarities with more complicated integrable systems,  is special in this context as its rather simple structure has allowed for a flurry of works studying in great detail various aspects of the model, ranging from analytical expressions to nonequilibrium steady-states \cite{prosen2016integrability,klobas2019time,klobas2020matrix,PhysRevE.102.062107,klobas2021exact}, operator and entanglement spreading \cite{gopalakrishnan2018operator,gopalakrishnan2018hydrodynamics,PhysRevLett.122.250603, klobas2021entanglement},  hydrodynamics in the context of Generalized Hydrodynamics (GHD) ~\cite{gopalakrishnan2018hydrodynamics,PhysRevLett.123.170603,buvca2021rule}, and thermalization properties~\cite{PhysRevLett.126.160602}.  Given these well-established properties,  it seems natural to use Rule 54 as a starting point in order to study more complex dynamics. In this sense, in this work we address the fate of the dynamics in Rule 54 when solitons are allowed to back scatter $R \leftrightarrow L$ at a given rate, thereby breaking the integrability of the model. 

Integrability breaking is a topic that also has a long history~\cite{PhysRevB.53.983,PhysRevB.76.245108,PhysRevLett.110.070602,PhysRevB.91.115130,PhysRevB.90.094417,PhysRevB.93.205121,PhysRevLett.115.180601,PhysRevB.94.245117,PhysRevLett.85.1092,PhysRevLett.103.216602,PhysRevB.83.035115,PhysRevB.88.115126,PhysRevLett.103.096402,PhysRevE.86.031122,PhysRevE.88.012108,stark2013kinetic,vidmar2016generalized,d2016quantum,PhysRevLett.125.180605,PhysRevB.102.184304,znidaric2021less, bulchandani2021onset}. The qualitative effects of integrability breaking perturbations are clear: even a small integrability-breaking perturbation leads to thermalization and diffusive transport at sufficiently long times~\cite{bertini2021finite}, quasiparticles acquire a finite lifetime, and Drude weights in a.c.~conductivities are broadened into Lorentzians. Yet making this picture more quantitative has only started to be possible very recently, in great part thanks to the advent of GHD~\cite{PhysRevX.6.041065,PhysRevLett.117.207201,doyon2020lecture}. This framework has allowed for a more refined picture. For instance, in some cases, the diffusion constant of the residual conserved quantities can be expressed purely in terms of \textit{hydrodynamic data} of the integrable model, through the rates at which nonconserved charges decay~\cite{friedman2020diffusive,durnin2020non}. Other recent results include \cite{PhysRevB.103.L060302,PhysRevLett.126.090602,PhysRevB.102.161110} --- see also \cite{bastianello2021hydrodynamics} for a recent review on progress on the physics of integrability breaking from the point of view of GHD.  The main bottleneck for a full understanding of the physics of integrability breaking is a lack of tools to compute the various decay rates governing the relaxation of nonconserved charges, which should be controlled by Fermi Golden's Rule (FGR). The matrix elements (``form factors'') of the integrability-breaking perturbation are usually out of reach analytically, especially for physical processes that involve high momentum transfer, and thus fall outside of  current hydrodynamic results.

 It is therefore crucial to find simple, tractable yet interacting models to study the physics of integrability breaking.  Understanding in detail such dynamics in a controlled setup would be of paramount importance as many nonequilibrium one dimensional quantum systems of experimental relevance are affected by perturbations that weakly break some conservation laws and a full account of these effects can affect the dynamics on accessible timescales~\cite{PhysRevX.8.021030,PhysRevLett.126.090602}.
 
  In this work, we address the fate of integrable dynamics in the presence of backscattering noise in the Rule 54 CA.  More explicitly, we consider a specific mechanism whereby a right mover turns into a left mover, and viceversa, with some small probability $p$ per time step (this will break the conservation of the chirality imbalance `$R-L$', while preserving the total number of solitons `$R+L$').  The resulting dynamics consists of the only remaining conserved charge, the density of particles of both chiralities in the system,  spreading diffusively, while the nonconserved charges, including the chirality imbalance, decay exponentially at long enough times. We focus on the dynamics of a tracer soliton and show that the resulting dynamics is chaotic with a characteristic \textit{self-diffusion} constant that depends on the perturbation strength and the density of particles in the system (for small enough $p$, so that the system remains near the  local equilibrium of the integrable limit at all times). This self-diffusion constant is generically different from the diffusion constant governing transport, which evades a full analytical treatment. The lack of a dephasing mechanism in this model (and in similar classical models) invalidates a perturbative treatment in the spirit of FGR as terms higher order in the perturbation strength grow in time as well (so that one cannot restrict to the lowest nontrivial order in perturbation theory). 

The remainder of this article is divided as follows. We start off in Section \ref{Sec: FFA} describing the model, its interpretation as a quantum circuit, and the observables of interest.  These correspond to the density of right and left moving solitons and their associated currents.  In Section \ref{Sec: GHD_formulas} we discuss the thermodynamics of Rule 54, and its hydrodynamics using GHD framework. We present the relevant GHD equations which take a very simple form as a result of the presence of just two types of solitons. The simplicity of the model allows us to also compute transport coefficients analytically. We next verify all these predictions (hydrodynamic equations + transport coefficients) using tensor network based numerical simulations finding excellent agreement. The next section, Section \ref{sec_int_breaking} deals with the breaking of integrability in the model.  Our main focus here is on tracer dynamics which one can understand in terms of a random walk with a dressed mean-free path as a result of a finite density of particles. Our predictions here are checked against numerical simulations. We also discuss transport from the point of view of linear response and argue that the FGR fails to be applicable in this system (in the sense that the leading nontrivial order in time dependent perturbation theory does not determine the decay rate of the nonconserved charges), a direct consequence of a lack of a dephasing mechanism in the system due to its classical nature. We close drawing some conclusions in Section \ref{sec_conclusion}.

\section{Brief description of the Rule 54 model} \label{Sec: FFA}
\subsection{Dynamical rule and states}
A cellular automaton (CA)\cite{martin1984algebraic} is an array of sites each  taking elements in a Boolean set $\mathcal{F}$, in our case $\mathcal{F}=\{ \ws, \bss\}$, with e.g. $\ws \cong 1$ and $\bss \cong 0$.  A \textit{state} is given by a specific configuration of these elements. States are evolved according to some dynamical rules that \textit{flip} an element, that is, $\ws \leftrightarrow \bss$, depending on the elements of its immediate neighbors. Given some state at time step $t$, $\underline{s}^{(t)}$, the Rule 54 flips an element if \textit{any} of its two adjacent sites takes the element $\ws$.  Labeling each site index by an element in $\mathbb{Z}/2$, this dynamical rule is implemented via the two step process
\begin{equation} \label{Eq: dyn_rule}
s_{i}^{(t+1/2)}=W[s_{i-1/2}^{(t)},s_{i}^{(t)},s_{i+1/2}^{(t)}], \hspace{0.1in} s_{i+1/2}^{(t+1)}=W[s_{i}^{(t+1/2)},s_{i+1/2}^{(t+1/2)},s_{i+1}^{(t+1/2)}],
\end{equation} 
with $i\in \mathbb{Z}$, $s_i \in \mathcal{F}$ and 
\begin{equation} \label{Eq: W_map}
W[s_a, s_b,s_c]\equiv s_a+s_{b}+s_{c}+s_as_{c} \hspace{0.1in} (\rm mod \hspace{0.1in} 2).
\end{equation}
We work in a system of $2L$ sites and periodic boundary conditions (pbc). Site indices labeled by an integer are referred to as being $A$ type, the rest being $B$ type.  A unit cell consists of an adjacent pair of $A$ and $B$ sites.  For convenience of notation we will only indicate the location of the unit cell by labeling the $i$'th site of $A$ type. E.g. for a system of 6 sites a state could be $\underline{s}=\us{1}{\ws\bss}\s\us{2}{\ws\ws}\s\us{3}{\bss\ws}$, where also for clarity of notation we will separate adjacent unit cells.  From now on we will only label the necessary unit cells.  
\subsection{Rule 54 as a unitary quantum circuit}
The Rule 54 CA has an interpretation in terms of a unitary circuit with local gates (with the prescription $\ws \cong \uparrow$ and $\bss \cong \downarrow$)
\begin{equation} 
\hat{W}_j=\hat{\sigma}_j^x(\mathbb{1}-\hat{d}_{j-1/2}\hat{d}_{j+1/2})+\hat{d}_{j-1/2}\hat{d}_{j+1/2},  \hspace{0.1in} j\in \mathbb{Z}/2, 
\end{equation} 
where $\hat{d}_j=\frac{1}{2}(\hat{\mathbb{1}}_j-\hat{\sigma}_j^z)=|\bss\rangle\langle\bss|_j$ and $\hat{\sigma}_j^a$, $a\in \{x,y,z\}$, are the Pauli matrices.  Each time step is then given by a two step cycle as
\begin{equation} \label{Eq: F_gates}
\hat{F}=\hat{W}_B\hat{W}_A, \hspace{0.1in} \hat{W}_{A/B}=\bigotimes_{j\in A/B} \hat{W}_j .
\end{equation}
Each unitary gate thus comprises three sites.  Note that, despite the \textit{classical} nature of the dynamics in the $z$ basis, the resulting dynamics generates operator entanglement~\cite{gopalakrishnan2018operator,PhysRevLett.122.250603} (as can be checked by evolving any local operator by (\ref{Eq: F_gates})).  In this paper we will make use of both the classical description of the model using large scale Monte Carlo (MC) simulations to access the long time dynamics as well as its quantum version through time-dependent matrix product operator (MPO) techniques~\cite{schollwock2011density} (for which we will take full advantage of the resulting noise-free dynamics to have access to precise values of transport coefficients already accessible at early to intermediate times).

\subsection{Observables}

\begin{figure}
\centering 
\includegraphics[scale=0.7]{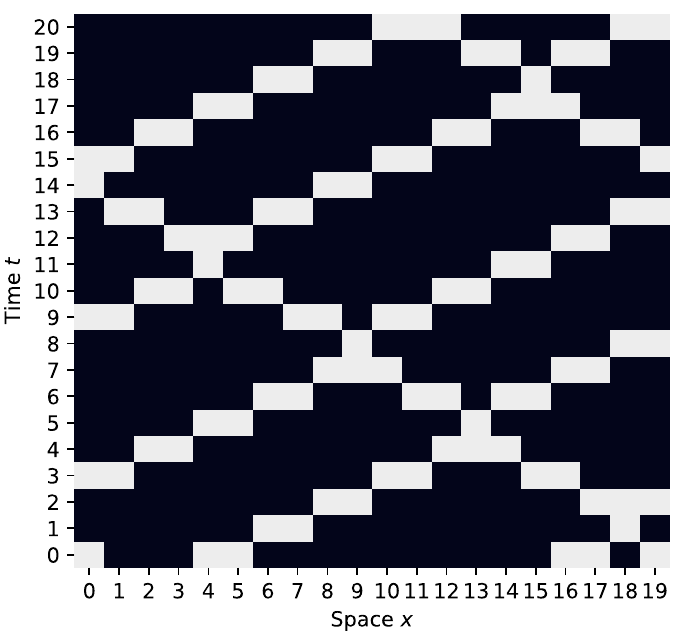}
\caption{Illustrative snapshot on the Rule 54 dynamics.  White (black) squares $\ws$ ($\bss$) take the value $1$ ($0$). }
\label{Fig: snapshot}
\end{figure}

Although the dynamical rules (\ref{Eq: W_map}) are hard to fit in the usual framework of Yang-Baxter integrability, they are believed to lead to integrable dynamics~\cite{1993CMaPh.158..127B,PhysRevLett.123.170603,buvca2021rule}. There are infinitely many conserved quantities, but we will only be concerned with the conservation of right and left moving solitons --- see e.g. Fig. \ref{Fig: snapshot}. These conservation laws will allow for a kinetic description of the system upon \textit{coarse graining} and whose discussion is postponed to the next Section. The number of \textit{right-movers} $N_R$ and \textit{left-movers} $N_L$ are independently conserved throughout the entire evolution, s.t. at any time step $t$ we have $N_R^{(t+1)}=N_R^{(t)}$ and $N_L^{(t+1)}=N_L^{(t)}$.  One can parameterize these in terms of the local \textit{observables} $\hat{\rho}_{R/L,x/x+1/2}$ s.t. $\hat{N}_{R/L}=\sum_i \hat{\rho}_{R/L,x/x+1/2}$ where 
\begin{eqnarray}\label{Eq: rho_R}
\hat{\rho}_{R,x}&=\us{$x$}{\ws\ws}+\bss\s\us{$x$}{\ws\bss}+\us{$x$}{\bss\ws}\s\bss,\\
\label{Eq: rho_L}
\hat{\rho}_{L,x+1/2}&=\us{$x$}{\ess\ws}\s\ws+\us{$x$}{\bss\ws}\s\bss+\us{$x$}{\ess\bss}\s\ws\bss,
\end{eqnarray}
where an empty square at site $x$ is the identity element, e.g. $\us{$x$}{\ess\ws}=\us{$x$}{\bss\ws}+\us{$x$}{\ws\ws}$, acting as diagonal operators in the computational basis. The two last elements in (\ref{Eq: rho_R}) and (\ref{Eq: rho_L}) come from a counting of right and left solitons during a collision event - see Fig.  \ref{Fig: snapshot}.  In the language of quantum operators, these conservation laws translate into $[\hat{F},\hat{N}_{R/L}]=0$. The conserved number of right and left movers gives rise to microscopic currents. Focusing on the right-movers first note that $\hat{\rho}_{R,x}^{(1)}-\hat{\rho}_{R,x}=-\hat{j}_{R,x+1/2}+\hat{j}_{R,x-1/2}$ where $\hat{j}_{R,x+1/2}=\us{$x$}{\ess\ws}\s\bss$.  By linearity, this implies 
\begin{equation} \label{eq_cont_R_lattice}
  \hat{\rho}_{R,x}^{(t+1)}-\hat{\rho}_{R,x}^{(t)}=\hat{j}_{R,x-1/2}^{(t)}-\hat{j}_{R,x+1/2}^{(t)}.
\end{equation}
Proceeding identically for the left-movers we get
\begin{eqnarray} \label{eq_cont_L_lattice}
\hat{\rho}_{L,x+1/2}^{(t+1)}-\hat{\rho}_{L,x+1/2}^{(t)}+\hat{j}_{L,x+1}^{(t)}-\hat{j}_{L,x}^{(t)}=0,
\end{eqnarray}
with the local current $\hat{j}_{L,x}=-\us{$x$}{\bss\ws}$. Note that there exists some gauge freedom in the way we choose the local currents as we can always add a local gradient term that would still give rise to the conservation of $N_{R/L}$.  An alternative basis to the $R/L$ basis that will be useful later is given in terms of the \textit{density of particles} (or density of  \textit{positive} movers) $\hat{\rho}_+=\hat{\rho}_R+\hat{\rho}_L$,  and \textit{imbalance} in the number of right and left movers (or density of \textit{negative} movers) $\hat{\rho}_-=\hat{\rho}_R-\hat{\rho}_L$.  (As discussed in the Introduction, the integrability breaking mechanism we will consider below will break the conservation of $\hat{\rho}_-$).

\section{GHD equations and transport coefficients in the Rule 54 model} \label{Sec: GHD_formulas}
\subsection{Thermodynamics}
The thermodynamics of the Rule 54 model was already discussed in \cite{gopalakrishnan2018hydrodynamics}. Equilibrium states are expressed in terms of the classical, one-dimensional partition function $Z=\sum_{\{\underline{s} \}}e^{-\mu_R N_R(\underline{s})-\mu_L N_L(\underline{s})}$ which can be computed using the transfer matrix 
\begin{eqnarray} \label{Eq: Transfer_matrix}
T=\left(\matrix{1&1&e^{-\mu_R-\mu_L}&e^{-\mu_R/2}\cr 
e^{-\mu_R-\mu_L}&e^{-\mu_R-\mu_L}&e^{-\mu_L}&e^{-\mu_R/2-\mu_L}\cr 
1&1&e^{-\mu_R-\mu_L}&e^{-\mu_R/2}\cr 
e^{-\mu_R/2}&e^{-\mu_R/2}&e^{-\mu_R/2-\mu_L}&e^{-\mu_R-\mu_L} }\right),
\end{eqnarray}
written in the $\{\bss\bss,\ws\bss,\bss\ws,\ws\ws \}$ basis.  Because Rule 54 is also integrable,  equilibrium states can also be formulated in terms of Bethe and Thermodynamic Bethe Ansatz (TBA) equations~\cite{yang1969thermodynamics, takahashi2005thermodynamics}.  The simplicity of the model permits us to write down a closed form for these~\cite{PhysRevLett.123.170603}. We can either work in the $R,L$ or the $+,-$ basis, being both related via 
$\vec{\rho}=\left( \matrix{ \rho_R, & \rho_L } \right)=\frac{1}{2}\mathbf{O} \left( \matrix{ \rho_+, & \rho_- } \right)$ with 
\begin{eqnarray}
\mathbf{O}=\left(
\matrix{
1 & 1\cr 1 & -1
}
\right).
\end{eqnarray}
The Bethe equations are expressed in terms of the {\em scattering kernel}, written in the $R,L$ basis 
\begin{eqnarray}
\mathbf{K}=\left(
\matrix{
1 & -1\cr -1 & 1
}
\right).
\end{eqnarray}
In the $\{+,-\}$ basis this matrix is diagonal with $0$ in the $+,+$ component. That is, $\rho_+$ is a zero mode of $\mathbf{K}$ (this will have important consequences for the transport properties of these modes, as we discuss next). It is useful to introduce the \textit{dressing} operation $\vec{h}^{\rm dr}=(\mathbb{1}+\mathbf{K}\mathbf{n})^{-1}\vec{h}$, for any vector $\vec{h} \in \mathbb{R}^2$.  (In the theory of integrable systems, this operation plays a crucial role, capturing the effects of interactions on physical quantities). The Bethe equation in Rule 54 is given by
\begin{equation} \label{Eq: Bethe}
\vec{\rho}=\mathbf{n}\vec{1}^{\rm dr},
\end{equation}
where $\vec{1}=\left( \matrix{ 1, & 1 } \right)$ and the Fermi factors in matrix form read $\mathbf{n}=\rm diag\left( \matrix{ n_R, & n_L } \right)$ with $n_{R/L}=\frac{\rho_{R/L}}{\rho_{R/L}^{\rm tot}}$, where $\rho_{R/L}^{\rm tot}$ total density of states.
The Bethe equation reads~\cite{PhysRevLett.123.170603}
\begin{eqnarray}
\rho_{R/L}^{\rm tot}=1+\rho_{L/R}-\rho_{R/L}.
\end{eqnarray}
 In a generalized Gibbs ensemble (GGE) with chemical potentials $\mu_R$ and $\mu_L$, the Fermi factors are given by $n_{R/L}=1/(1+\exp(\epsilon_{R/L}))$, where the quantities $\epsilon_{R/L}$ are called \textit{pseudoenergies} of the quasiparticles. They fulfill the TBA equations~\cite{PhysRevLett.123.170603}
\begin{equation} \label{Eq: TBA}
\epsilon_{R/L}=\mu_{R/L}+\log \frac{1+e^{-\epsilon_{R/L}}}{1+e^{-\epsilon_{L/R}}}.
\end{equation} 
Computing thermodynamic quantities through solving those equations or using the transfer matrix~(\ref{Eq: Transfer_matrix}) give identical results. 

\subsection{Hydrodynamics} \label{sec_hydro}
Hydrodynamics is a theory based upon \textit{coarse graining}, and the assumption of local equilibrium.  Coarse graining is the procedure by which we consider the dynamics at a \textit{fluid cell} level and in the limit of \textit{sufficiently} large space-time points $x,t$ \cite{PhysRevX.6.041065,PhysRevLett.117.207201,doyon2020lecture,de2021correlation}. The exact mechanism through which this happens is via averaging Heisenberg operators $\hat{\mathcal{O}}_x^{(t)}$ w.r.t.~a Gibbs ensemble of locally conserved charges - that is, we assume that the system is in \textit{local equilibrium} 
\begin{equation}
\langle \hat{\mathcal{O}}_x^{(t)}\rangle_{\mu_R,\mu_L} \equiv \frac{1}{Z}\tr[\hat{\mathcal{O}}_x^{(t)} e^{-\sum_x \mu_R(x)\hat{\rho}_{R,x}+\mu_L(x)\hat{\rho}_{L,x}}].
\end{equation}
%An alternative view to this average that will be helpful later on is using MC sampling 
%\begin{equation}
 %   \langle \hat{\mathcal{O}}_x^{(t)}\rangle_{\mu_R,\mu_L} = \sum_{\alpha} \omega(\alpha) \underline{s}_\alpha \cdot \hat{\mathcal{O}}_x^{(t)} \underline{s}_\alpha = \sum_\alpha \omega(\alpha) \underline{s}_\alpha^{(t)} \cdot \hat{\mathcal{O}}_x \underline{s}_\alpha^{(t)}, 
%\end{equation}
%where $\alpha$ labels each of the possible states labeled by the chemical potentials $\mu_R$, $\mu_L$, and $\omega (\alpha)$ is a probability measure for each configuration labeled by $\alpha$. 
To simplify the notations, we will simply denote such expectation values as $\mathcal{O}_x^{(t)} \equiv  \langle \hat{\mathcal{O}}_x^{(t)}\rangle_{\mu_R,\mu_L}$ making implicit assumption that such quantities are evaluated w.r.t. an arbitrary background state.  

 Hydrodynamics dictates that after \textit{local equilibration} has occured,  the exact (lattice) relations $\hat{\rho}^{(t+1)}_{R/L,x}-\hat{\rho}^{(t)}_{R/L,x}+\hat{j}^{(t)}_{R/L,x+1}-\hat{j}^{(t)}_{R/L,x}=0$ may be replaced by the coarse grained ones $\partial_t \rho_{R/L,x}^{(t)} + \partial_x j_{R/L,x}^{(t)}=0$. 
So far we have focused our analysis on the two local conservation laws corresponding to $R/L$ movers (alternatively $+/-$ movers).  Integrable systems have infinitely many local conserved laws. In Rule 54 those infinitely many remaining conservation laws correspond to \textit{adjacent spacings} of solitons of the same chirality \cite{gopalakrishnan2018operator}, but do not play any role on the transport properties of the quantities of interest. 

Having argued the emergence of two conservation laws in the model at hand, let us now see how the GHD framework allows us to go further and compute transport coefficients.  The continuity equations were worked out already in Ref. \cite{gopalakrishnan2018hydrodynamics}. These read 
\begin{equation} \label{Eq: continuity_plusminus}
\partial_t \rho_{+/-} + \partial_x j_{+/-} = 0, 
\end{equation}
where the currents are given by $j_{+/-}= \rho_Rv_R \pm \rho_Lv_L$.  We emphasize that Eq. (\ref{Eq: continuity_plusminus}), as opposed to Eqs. (\ref{eq_cont_R_lattice}, \ref{eq_cont_L_lattice}), is emergent, and assumed to hold at the Euler scale $x \sim t$ \cite{PhysRevX.6.041065,PhysRevLett.117.207201}. To find the velocities $v_{R/L}$ we proceed as in \cite{gopalakrishnan2018hydrodynamics}. Consider a right-mover starting at $(x^0,t^0)=(0,0)$ and ending at $(x,t)$. As it travels to the right with a bare velocity $v_R^0=+1$ it will encounter left-movers that started closer than $x-v_Lt$ causing time-delays so that after $t$ time steps $t=x+\rho_L(x-v_Lt)$. This gives $v_R=x/t=(1+\rho_Lv_L)/(1+\rho_L)$. Similar arguments for a left-mover leads to $v_L=(-1+\rho_Rv_R)/(1+\rho_R)$. Solving these two equations gives 
\begin{equation} \label{Eq: velocities}
v_{R/L}=\pm 1 \mp \frac{2\rho_{L/R}}{1+\rho_R+\rho_L}.
\end{equation}
As a result, we have in particular $j_+=\rho_-$.  This equation holds microscopically, which means that this Euler relation is exact (we do not have higher order diffusive corrections). To see this we start from the definition of $\hat{\rho}_{-,x+1/2}=\hat{\rho}_{R,x}-\hat{\rho}_{L,x+1/2}$, and using Eqs. (\ref{Eq: rho_R}-\ref{Eq: rho_L}) one finds $\hat{\rho}_{-,x+1/2}= \hat{j}_{+,x} +\mathcal{O}(\Delta \hat{O}_x)$, that is, the equality holds up to derivative (gauge) terms of the form $\Delta\hat{O}_x \equiv \hat{O}_{x+1}-\hat{O}_x$; see the remark after Eq. (\ref{eq_cont_L_lattice}). The current for $\rho_-$ is more complicated and includes diffusive corrections.  The Bethe equation (\ref{Eq: Bethe}) gives us
\begin{equation} \label{Eq: rho_v}
\rho\vec{v}=\mathbf{n}\vec{v}^{\rm dr},
\end{equation}
where $\vec{v}=\left( \matrix{ v_R, & v_L }\right)$ , and $\rho=\rm diag \left( \matrix{\rho_R, & \rho_L}\right)$. Differentiating w.r.t. $t$ and $x$ Eqs. (\ref{Eq: Bethe}) and (\ref{Eq: rho_v}) and after some straightforward algebra gives the following advection equation written in terms of Fermi factors \cite{PhysRevX.6.041065,PhysRevLett.117.207201}
\begin{equation}
\partial_t \vec{n}+\mathbf{v}\partial_x\vec{n}=0.
\end{equation}
\subsection{Transport coefficients} \label{Sec: transport}
The Euler equations (\ref{Eq: continuity_plusminus}) admit diffusive corrections and they give rise to non-zero transport coefficients. We are interested in the d.c. conductivity and Drude weights.  Within GHD these transport coefficients can be computed exactly in Rule 54. First we spell out explicitly the quantities of interest. The conductivity tensor reads \cite{bertini2021finite,PhysRevLett.119.110603}
\begin{equation} \label{Eq: dc_conductivity}
\sigma_{a,b}(\omega)=\frac{1}{2}G_{a,b}(0)+\sum_{t=1}^\infty G_{a,b}(t)e^{i\omega t},
\end{equation}
with the connected current-current correlation function $G_{a,b}(t)=\frac{1}{L}\langle \hat{J}^{(t)}_a\hat{J}^{(0)}_b\rangle^c \equiv \frac{1}{L}(\langle \hat{J}^{(t)}_a\hat{J}^{(0)}_b\rangle - \langle \hat{J}^{(t)}_a\rangle \langle \hat{J}^{(0)}_b\rangle)$ where in our case we will choose the basis of density and imbalance of particles, so that $a,b\in \{+,-\}$.  Sending $t\to \infty$ in $G_{a,b}(t)$ we obtain the Drude weight 
\begin{equation}
    D_{a,b}=\lim_{t\to\infty}G_{a,b}(t).
\end{equation}  
Taking instead the zero-frequency (long-wavelength) limit of (\ref{Eq: dc_conductivity}) we obtain the d.c. conductivity 
\begin{equation} \label{Eq: dc_time}
\sigma_{a,b}^{\rm d.c.}=\frac{1}{2}G_{a,b}(0)+\sum_{t>0}G_{a,b}(t)\geq 0,
\end{equation} 
where the last inequality follows from the known fact that the integrated autocorrelation function is always non-negative (in particular, the imaginary part of the d.c. conductivity vanishes since the conductivity (\ref{Eq: dc_conductivity}) is an odd function of frequency \cite{bertini2021finite}). Of course, for integrable systems this quantity often diverges. It is then customary to split the d.c.~conductivity in terms of a regular and a divergent component, the latter being proportional to the Drude weight. From now on we will refer to this regular part as the d.c.~conductivity, i.e. 
\begin{equation} \label{Eq: dc_cond_redef}
    \sigma_{a,b}^{\rm d.c.}\equiv \frac{1}{2}\tilde{G}_{a,b}(0)+\sum_{t>0}\tilde{G}_{a,b}(t),
\end{equation}
with $\tilde{G}_{a,b}(t)\equiv G_{a,b}(t)-D_{a,b}$. To compute these two quantities we resort to the GHD formalism (for recent reviews on the GHD formalism and its connection with hydrodynamic matrices see \cite{de2019diffusion}). Note that, using the Bethe and TBA equations (\ref{Eq: Bethe}) and (\ref{Eq: TBA}) allows us to switch between the thermodynamic variables $\{ \mu_R, \mu_L \} \leftrightarrow \{ \epsilon_R, \epsilon_L \} \leftrightarrow \{ \rho_R, \rho_L \} \leftrightarrow \{ n_R, n_L \}$. The Drude matrix reads \cite{PhysRevB.96.081118,SciPostPhys.3.6.039,de2019diffusion}
\begin{equation} \label{eq_Drude}
\mathbf{D}=(\mathbb{1}+\mathbf{n}\mathbf{K})^{-1}\rho(\mathbb{1}-\mathbf{n})\mathbf{v}^2(\mathbb{1}+\mathbf{K}\mathbf{n})^{-1}
\end{equation}
from where we extract $D_{a,b}$ for a given equilibrium state determined by $\mu_{R,L}$. To simplify the expressions of the Drude matrix let us consider $\mu_R=\mu_L$. The Drude matrix has components
\begin{equation} \label{Eq: Drude_--}
D_{+,+}=2\frac{\rho_L(1-\rho_L)}{(1+2\rho_L)^2}, \hspace{0.1in} D_{-,-}=2\frac{\rho_L(1-\rho_L)}{(1+2\rho_L)^4},
\end{equation}
the others vanishing. To get the d.c. conductivity matrix we first need the susceptibility matrix \cite{SciPostPhys.3.6.039,de2019diffusion}
\begin{equation}
\mathbf{C}=(\mathbb{1}+\mathbf{n}\mathbf{K})^{-1}\rho(\mathbb{1}-\mathbf{n})(\mathbb{1}+\mathbf{K}\mathbf{n})^{-1}.
\end{equation}
We will also need the diffusion kernel, whose components are \cite{PhysRevLett.121.160603,de2019diffusion}
\numparts
\begin{eqnarray} \label{eq_diagonal_D}
\tilde{\mathcal{D}}_{a,a}&= \frac{1}{2}\rho_b(1-n_b)\left( \frac{K^{\rm dr}_{a,b}}{\rho^{\rm tot}_a}\right)^2|v_a-v_b|,\\
\tilde{\mathcal{D}}_{a,b\neq a}&=\frac{1}{2}\rho_a(1-n_a)\left( \frac{K^{\rm dr}_{a,b}}{\rho^{\rm tot}_a}\right)^2|v_a-v_b|=\left(\frac{\rho_b^{\rm tot}}{\rho_a^{\rm tot}}\right)^2\tilde{\mathcal{D}}_{b,b}.
\end{eqnarray}
\endnumparts
The diagonal elements of this matrix can be seen as the variance of the quasiparticle fluctuations and were found independently using different approaches in \cite{PhysRevLett.121.160603,gopalakrishnan2018hydrodynamics}. The off-diagonal ones can be viewed as arising from quasiparticle scatterings \cite{PhysRevLett.121.160603,de2019diffusion}. From here one obtains the diffusion matrix \cite{PhysRevLett.121.160603,de2019diffusion}
\begin{equation} 
\mathcal{D}=(\mathbb{1}+\mathbf{n}\mathbf{K})^{-1}\rho^{\rm tot}\tilde{\mathcal{D}}(\rho^{\rm tot})^{-1}(\mathbb{1}+\mathbf{n}\mathbf{K}).
\end{equation}
The diffusion matrix gives us the diffusive (or ``Navier-Stokes'') corrections to the hydrodynamic equation (\ref{Eq: continuity_plusminus}), valid at space-time scales $x\sim t^{1/2}$ and results in~\cite{PhysRevLett.121.160603}
\begin{equation} \label{Eq: N-S GHD}
\partial_t\vec{\rho}+\partial_x(\mathbf{v}[\rho]\vec{\rho})=\partial_x(\mathcal{D}[\rho]\partial_x\vec{\rho}).
\end{equation}
Written explicitly in terms of Fermi factors in the $R,L$ basis we get 
\numparts
\begin{eqnarray} \label{eqNavierStokesFermi}
\fl \partial_t n_R + \frac{1}{1+2 n_L}  \partial_x n_R - \frac{n_L(1-n_L)}{(1+2 n_L)^3} \partial_x^2 n_R + \frac{n_R(n_R-1) }{(1 + 2 n_R) (1 + 2 n_L)^2 } \partial_x^2 n_L &=0,  \\
\fl \partial_t n_L + \frac{1}{1+2 n_R}  \partial_x  n_L -  \frac{n_R(1-n_R)}{(1+2 n_R)^3} \partial_x^2 n_L + \frac{n_L(n_L-1) }{(1 + 2 n_L) (1 + 2 n_R)^2 } \partial_x^2 n_R &= 0.
\end{eqnarray}
\endnumparts
One can also check that indeed $\mathcal{D}_{+,+}=\mathcal{D}_{+,-}=0$, so that diffusive corrections for particle number vanish, as argued earlier. The d.c. conductivity matrix then reads 
\begin{equation}
\sigma^{\rm d.c.}=\mathcal{D}\mathcal{C}.
\end{equation}
(In the literature, this is also known as the Onsager matrix \cite{de2019diffusion}). We will test the prediction 
\begin{equation}
\sigma_{-,-}^{\rm d.c.}=4\frac{n_L(1-n_L)n_R(1-n_R)}{(1+2n_L)^2(1+n_R+n_L)(1+2n_R)^2},
\end{equation}
(which is valid for any $\mu_{R,L}$) the other components being zero. For completeness let us report the hydrodynamic equations within linear response (LR) at fixed background state $\rho^*$ with $\rho_-^*=0$ since this will be helpful when discussing integrability breaking in the model.  The hydrodynamic equations in this regime then read
\numparts
\begin{eqnarray}
\partial_t \delta \rho_+ +\partial_x \delta \rho_- &= 0, \label{Eq: rhop_eom}\\
\partial_t  \delta \rho_-+\partial_x j_-&=0,
\label{Eq: rhom_eom}
\end{eqnarray}
\endnumparts
where $\delta \rho_{+/-} = \rho_{+/-}-\rho_{+/-}^*$, and we have used $j_+=\delta \rho_-$. While the density of particles move purely ballistically,  the imbalance does so with diffusive corrections $j_-=\frac{1}{(1+2n)^2}\delta \rho_+ -2\frac{n(1-n)}{(1+2n)^3}\partial_x \delta \rho_-+...$,  where the Fermi factor $n$ is given now by $n=1/(1+e^\mu)=\rho_+^*/2$ ($\mu$ is the chemical potential associated to the background density of particles), and the $\cdots$ in $j_-$ correspond to higher order terms in a gradient expansion of the current (they can be disregarded). The terms kept in the expansion of $j_-$ instead correspond to the ballistic (Euler) and diffusive (Navier-Stokes) contributions.

\subsection{Testing GHD predictions}
\begin{figure}
\centering
\includegraphics[scale=0.35]{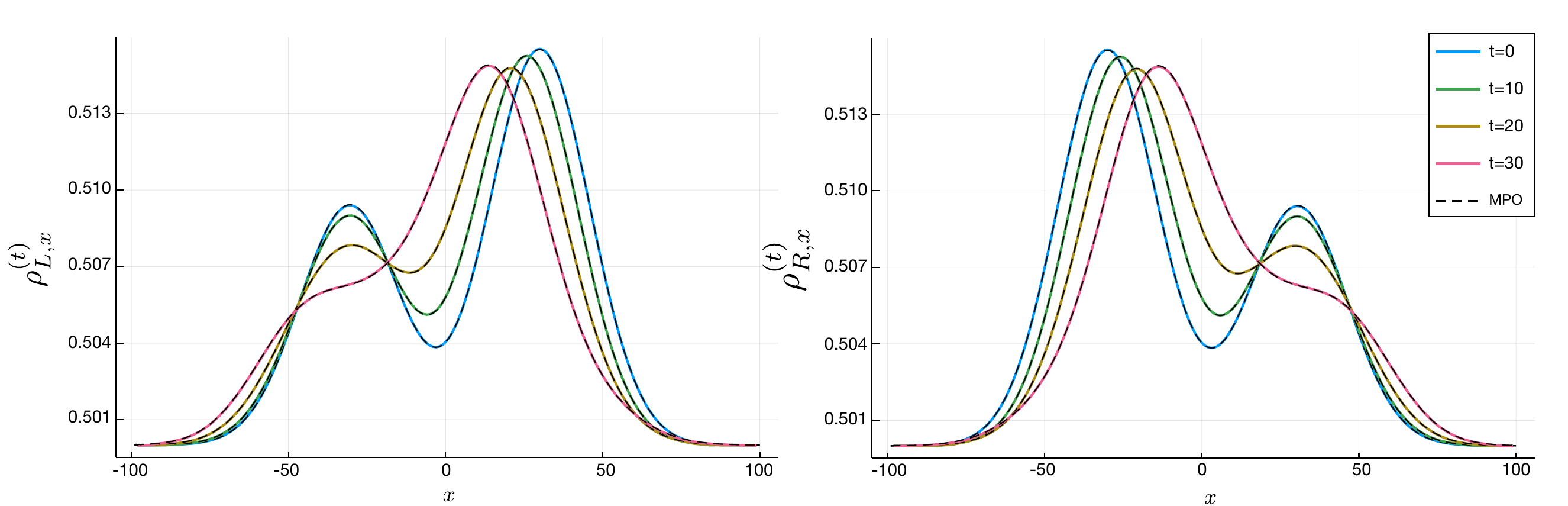}
\caption{\textbf{GHD predictions vs MPO: smooth initial states. } Left (right): left (right) movers density profiles as a function of space-time for an initial state given by $\mu_R(x)=\alpha\exp(-\beta(x+30)^2)$ and $\mu_L(x)=\alpha \exp(-\beta(x-30)^2)$ with $\alpha=0.1$ and $\beta=0.002$. MPO results using $\chi=32$.}
\label{Fig: GHD_vs_MPO_rhos}
\end{figure}

We verify the GHD predictions against first-principle numerical simulations.  Let us first comment on previous work verifying numerically the various GHD formulas for transport coefficients. The Drude weight formula -- corresponding to Eq. (\ref{eq_Drude}) in Rule 54, was first derived in \cite{SciPostPhys.3.6.039,PhysRevB.96.081118} within GHD by means of a form-factor expansion while retaining only one particle-hole excitations.  Prior to this, Drude weights had been computed in different integrable models by different means in {\it e.g.} \cite{fujimoto1998exact,zotos1999finite,klumper2002thermal,sakai2003non, PhysRevLett.106.217206, PhysRevLett.111.057203, PhysRevLett.119.110603}. First numerical evidence of the GHD formula for the Drude weight came in Refs. \cite{PhysRevB.97.045407,PhysRevLett.119.020602}.  Euler scale GHD equations have passed various sound numerical tests (and also experimental, see Ref.~\cite{bouchoule2021generalized} and references therein), using either a starting bipartition protocol \cite{PhysRevX.6.041065, PhysRevLett.117.207201}, as well as starting from arbitrary \textit{smooth} initial profiles \cite{PhysRevLett.119.220604,PhysRevB.97.045407,PhysRevLett.123.130602}.  Going beyond the Euler scale, diffusive corrections were incorporated within the GHD framework in Refs. \cite{PhysRevLett.121.160603,gopalakrishnan2018hydrodynamics}.  At variance with the derivation of Euler scale hydrodynamics within GHD, diffusive hydrodynamics necessarily incorporates two particle-hole excitations. The most direct application of \textit{diffusive} GHD has been within the context of spin transport in the XXZ spin chain~\cite{de2019diffusion,gopalakrishnan2019kinetic}. Here, by means of matrix product operator (MPO) techniques we compute expectation values as well as correlation functions of operators evolved under Heisenberg dynamics w.r.t.~spatially inhomogeneous mixed initial states.  We verify both GHD equations as well as transport coefficients (including the d.c. conductivity). For the technical details of the algorithm used we refer to Sec. \ref{Sec: MPO_calculations} of the Appendix. 

\begin{figure}
\centering
\includegraphics[scale=0.45]{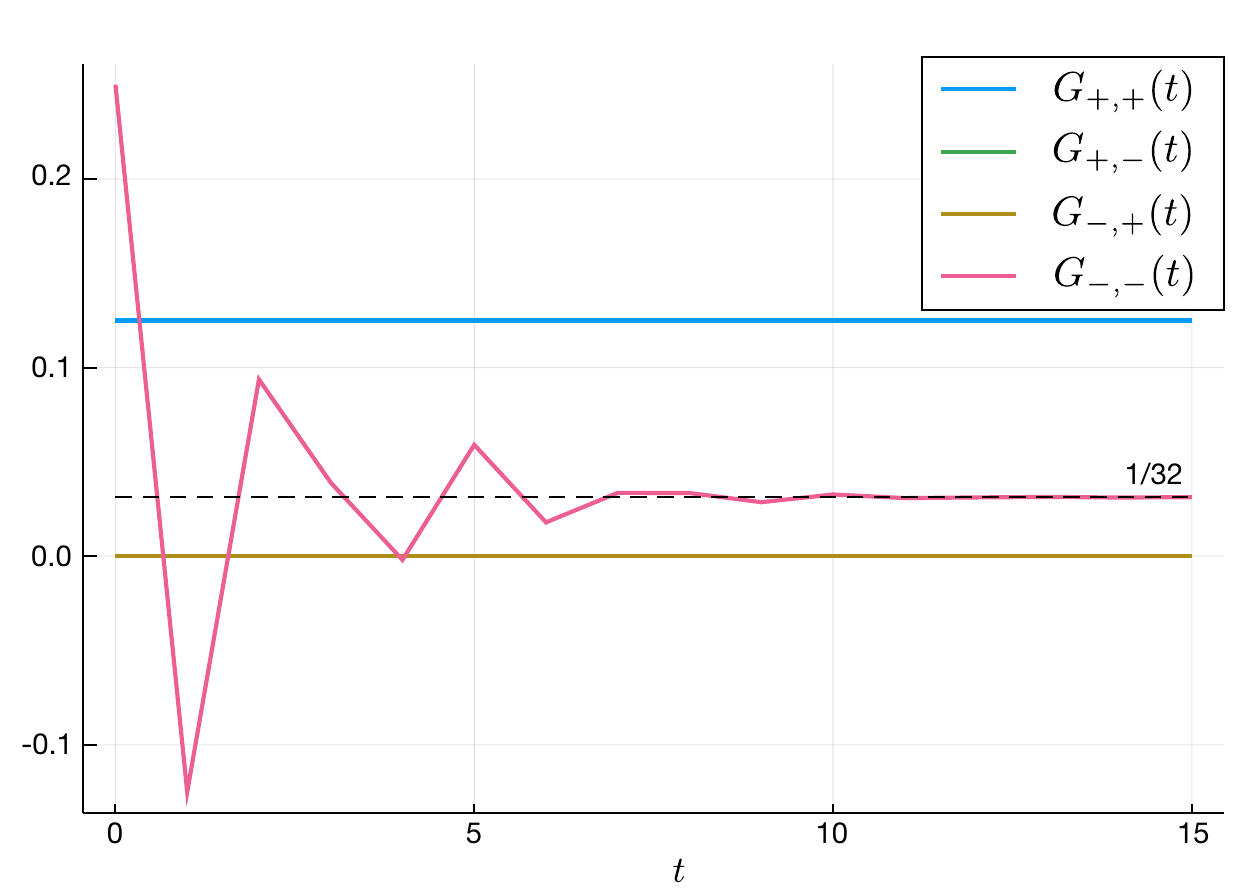}
\caption{\textbf{GHD predictions vs MPO: Kubo correlators.} Current-current correlation function $\frac{1}{L}\langle \hat{J}_a^{(t)}\hat{J}_b^{(0)}\rangle ^c$ for $a,b \in \{+,-\}$ at $\mu_R=\mu_L=0$ using MPOs of $\chi=128$ and predicted Drude weight $D(\mu=0) = 1/32$.} 
\label{Fig: GHD_vs_MPO_JJ}
\end{figure}

\begin{figure}
\centering 
\includegraphics[scale=0.35]{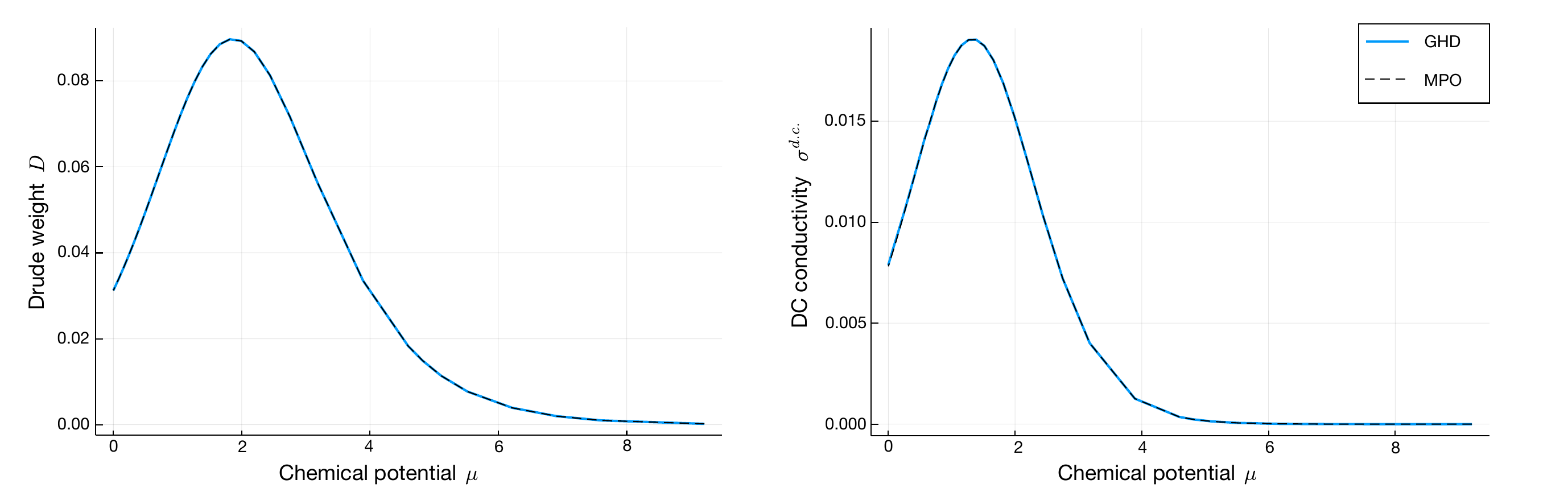}
\caption{\textbf{GHD predictions vs MPO: Drude weight and conductivity vs $\mu$.} Transport coefficients Drude weight (left) and d.c. conductivity (right) as a function of chemical potential $\mu_R=\mu_L=\mu$. MPO results using $\chi=128$.} 
\label{Fig: GHD_vs_MPO_transport}
\end{figure}
\begin{figure}
\centering
\includegraphics[scale=0.4]{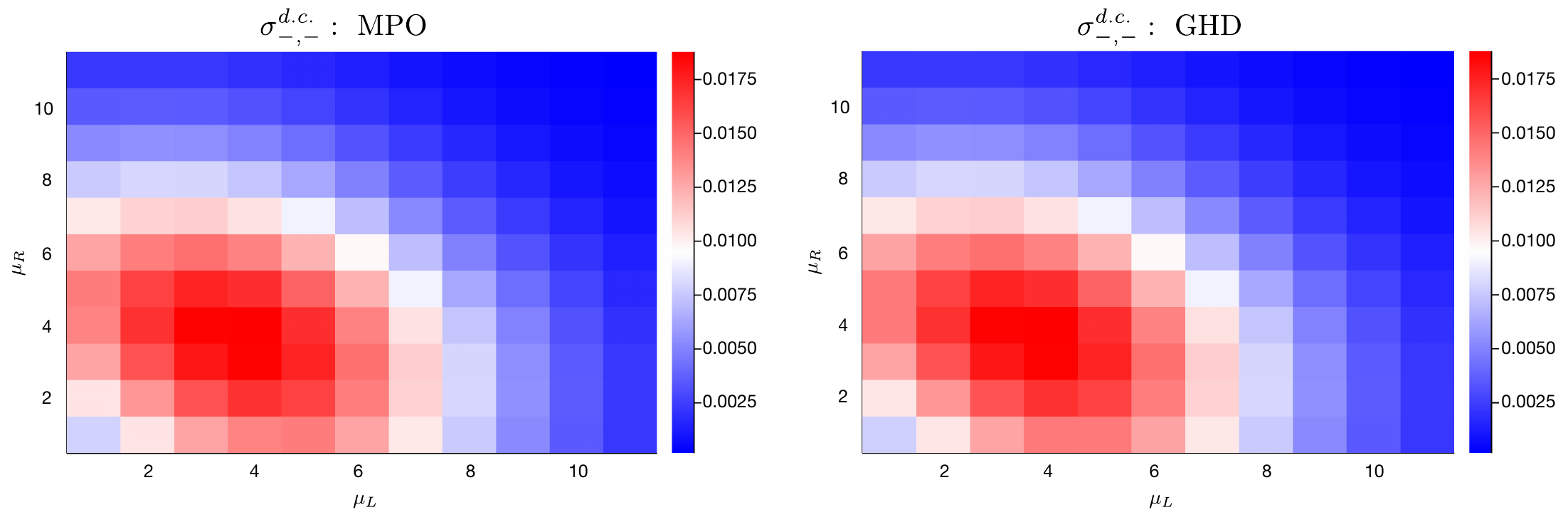}
\caption{\textbf{GHD predictions vs MPO: conductivity vs $\mu_{R/L}$. } Left (right): d.c. conductivity tensor $\sigma_{a,b}^{d.c.}$ vs. $\mu_{R/L}$ via MPO (GHD).}
\label{Fig: GHD_vs_MPO_dc_cond_heatmap}
\end{figure}

In Fig.  \ref{Fig: GHD_vs_MPO_rhos} we compare the GHD predictions solving (\ref{Eq: continuity_plusminus}) and the MPO based numerical results using a relatively small bond dimension $\chi=32$. We choose a system size of $L=200$ to allow for smooth initial conditions, which in this case we choose to be a superposition of gaussians in $\mu_R(x)$ and $\mu_L(x)$.  For the GHD equations, we first solve the TBA and Bethe equations.  This fixes the initial conditions $\rho_R(0,x)$ and $\rho_L(0,x)$. We then feed these initial conditions into (\ref{Eq: N-S GHD}), that we solve using the Crank-Nicholson algorithm. We remark that the diffusive corrections in (\ref{Eq: N-S GHD}) are so small that they do not make any difference on those plots. However, as we shall see such diffusive corrections can be better captured when studying instead transport via current-current correlation functions. We remark that the simplicity of the set-up allows for very fast computation times, both for the GHD and MPO results (of the order of seconds for solving the GHD equations and of minutes for the MPO time evolution on a regular laptop). 

In Fig. \ref{Fig: GHD_vs_MPO_JJ} we show the results of computing (connected) current-current correlation functions $\frac{1}{L}\langle \hat{J}_a^{(t)}\hat{J}_b^{(0)}\rangle^c$ using MPO time evolution when the background state is $\mu_R=\mu_L=0$.  Again, we find perfect agreement between our numerical results and the GHD predictions (\ref{Eq: Drude_--}).  Using (\ref{Eq: dc_time}) we verify as well the GHD predictions for the d.c. conductivity which in this case takes the value $\sigma^{d.c.}=1/128$.  

The case $\mu=0$ is special for the following reason. Any product of projectors $\ws$, $\bss$, has e.v.  for fixed $\mu$ of $\langle \hat{\mathcal{O}} \rangle_\mu = \frac{e^{a\mu}}{(1+e^{\mu})^b(3+e^{\mu})}$, with $a, b \in \mathbb{Z}^+_0$ . This implies that at $\mu=0$,  $\hat{j}_{-,x}$ evolves onto a linear combination of such product of operators with integer weight, i.e. $\hat{j}_{-,x}^{(t)}=\sum_i n_i \hat{\mathcal{O}}_i$, with $n_i$ some integer,  and so it follows $\frac{1}{L}\langle \hat{J}_-^{(t)}\hat{J}_-^{(0)}\rangle_{\mu=0} = \sum_i m_i/2^{b_i}$, with $m_i, b_i$ some positive integers. In fact it is easy to realize that $\frac{1}{L}\langle \hat{J}_-^{(t)}\hat{J}_-^{(0)}\rangle_{\mu=0}=n(t)/2^{2t+1}$ for $t>0$, where $n(t)$ is again a positive integer. The denominator in this expression comes from the fact that at any given time $t>0$, the longest string cannot have greater length than $2t+1$.  With this information and the numerical results (which are exact at least for big enough $\chi$) we can extract easily the values of $n(t)$ (a priori any floating point number from the numerical results could be interpreted as being either rational or irrational; this analysis discards the latter possibility).  It is quite remarkable that, despite the fact that $n(t)$ results in a sequence of odd numbers (e.g. $n(3)=5$,  $n(4)=-1$, $n(5)=121$, $n(10)=68667$, $n(12)=1045967$), they conspire to add to something very simple,  as reflected in the value of the d.c. conductivity $\sigma_{-,-}^{\rm d.c.}=1/128$.  

We conclude this section by benchmarking the GHD predictions for $D(\mu)$ and $\sigma^{\rm d.c.}(\mu)$ in Fig. \ref{Fig: GHD_vs_MPO_transport} finding again perfect agreement between predictions and MPO results. We remark that the $\sigma^{\rm d.c.}$ values are particularly small owing to the small diffusive corrections in Rule 54 and yet there is no appreciable discrepancy between the GHD predictions and those obtained using first-principle, \textit{microscopic} calculations based on MPO techniques. We extend the benchmarks of $\sigma^{\rm d.c.}$ to arbitrary $\mu_{R/L}$ in Fig. \ref{Fig: GHD_vs_MPO_dc_cond_heatmap}.

\section{Breaking integrability in the Rule 54} \label{sec_int_breaking}

\subsection{Interlude: integrability breaking in Hamiltonian systems} \label{Sec: Interlude}
Before embarking on breaking integrability in the Rule 54 model, it is worth taking a brief detour and discuss what we expect from the study of integrable Hamiltonian systems with at least one broken conservation law. This is a topic with a long history in the literature of integrable systems (see e.g.~\cite{PhysRevLett.111.197203,PhysRevB.84.054304,PhysRevB.94.245117,langen2016prethermalization,PhysRevB.95.104304,d2016quantum,vidmar2016generalized}), that has regained some interest recently in the context of GHD~\cite{friedman2020diffusive,durnin2020non,bastianello2021hydrodynamics,PhysRevB.102.161110,PhysRevB.103.L060302}. Our main goal is to review briefly the assumptions that go into this framework, as they appear to fail for Rule 54. 

Consider a system described by an integrable Hamiltonian $\hat{H}_0$ and a perturbation $g\hat{V}=g\int dx \hat{v}(x)$ that breaks the conservation of charge $\hat{q}_i$, so that the total Hamiltonian is given by $\hat{H}=\hat{H}_0+g\hat{V}$.  A necessary requirement for $\hat{V}$ to break the conservation of $\hat{Q}_i=\int dx \hat{q}_i(x)$ is that $[\hat{Q}_i,\hat{V}] \neq 0$. The dynamics of the e.v. of the charge, $q_i=\langle \hat{q}_i\rangle$, up to $\mathcal{O}(g^3)$ is governed by 
\begin{equation} \label{eq_2nd_order_pt}
\partial_t q_i=g^2\int_{-t}^{t}ds \langle [\hat{V}^0(s),\hat{Q}_i]v(0)\rangle^c + \mathcal{O}(g^3),
\end{equation}
where as usual when dealing with perturbation theory, operators are evolved in the interaction picture, $\hat{O}^0(t)=e^{i\hat{H}_0t}\hat{O}e^{-i\hat{H}_0t}$ and the only assumption so far is a background homogeneous state with density matrix $\rho_0=Z^{-1}\exp (-\sum_j \beta^j \hat{Q}_j)$ and $[\hat{Q}_j,\hat{Q}_k]=0$ for any pair of charges.  Note that neglecting $\mathcal{O}(g^3)$ terms and assuming a continuous spectrum of $\hat{H}_0$ we recover the Fermi Golden Rule (FGR) expression 
\begin{equation} \label{eq_FGR}
\partial_t q_i=4\pi^2g^2\sum_n \delta(\Delta \epsilon)\delta(\Delta p)\Delta q_i|\langle n| \hat{u}|\rho\rangle|^2,
\end{equation}
where an insertion of the identity in terms of eigenstates of $\hat{H}_0$ has been made, and e.v.s have been expressed in terms of the quasiparticle density $\rho$.   The rhs is sometimes referred to as the \textit{drift}. The terms $\Delta \epsilon$, $\Delta p$, are the difference in energy (w.r.t. the unperturbed Hamiltonian) and momentum, between states $\ket{n}$, $\ket{\rho}$, and $\Delta q_i$ is the difference in charge eigenvalue of $\hat{Q}_i$ in $\ket{n}$ vs. $\ket{\rho}$.  Lastly, we have implicitly taken $t \to \infty$, $L \to \infty$, to bring down the $\delta (\cdot)$ terms.  An analogous expression to (\ref{eq_FGR}) may be found if instead of an integrable Hamiltonian $H_0$, we started off from a Hamiltonian with at least one conserved charge $\hat{Q}_i$,   so this result is rather universal.  One key assumption for the validity of FGR in many-body systems, as pointed out in {\it e.g.} \cite{mallayya2019prethermalization}, is that the system should \textit{(i)} \textit{equilibrate quickly} and $\textit{(ii)}$ be \textit{weakly-coupled}.  The fast equilibration condition means that after a finite time $\tau^*$, the system relaxes to the diagonal equilibrium ensemble of the unperturbed Hamiltonian -- this ignores hydrodynamic tails effects.  The weak coupling condition means $g \tau^* \ll 1$.  (See \cite{mallayya2019prethermalization} for precise meaning of these conditions).
\subsection{Setup in the noisy Rule 54}
When considering breaking the conservation of either of the two charges in Rule 54, we find it more natural to break the conservation in the imbalance of solitons. From a physical standpoint this setup should mimic the physics of breaking the conservation of momentum in a Bose gas system in the presence of Galilean invariance (recall the imbalance corresponds to the current of density of solitons, $j_+=\rho_-$).   One way to implement this choice of integrability breaking is to convert a right mover into a left mover with probability $p$, and viceversa.  Ultimately we seek a r.h.s.  in (\ref{Eq: rhom_eom}) of the form $-\rho_-/\tau$ in the spirit of a relaxation time approximation \cite{lifschitz1983physical}, where the relaxation time $\tau \propto 1/p$. This would result in an exponential decay over time of the nonconserved charges $\rho_-$.  Microscopically,  the integrability breaking mechanism should be of the form $\us{x}{\ess \ws}\s\ws \rightleftarrows \us{x}{\ws\ws}$ with probability $p$ (collision terms appearing in $\hat{\rho}_{R/L,x/x+1/2}$ are left intact).  One can show that in order for the perturbation to preserve the number of particles it must act on a specific subspace. There are various choices for such a subspace which we report in \ref{sec_set_Kraus}, the simplest of which is given when applying the following projector
\begin{equation} \label{Eq: Pi}
\hat{\Pi}_x=\hat{\Pi}_{R,x}+\hat{\Pi}_{L,x+1/2},
\end{equation} 
where $\hat{\Pi}_{R,x} \equiv \ess \bss \s\us{x}{\ws\ws}\s\bss \bss $, $\hat{\Pi}_{L,x+1/2} \equiv \ess \bss \s\us{x}{\bss \ws}\s\ws\bss $. The dynamics is then given by a two step process: first, we evolve by one time step via the unitary map $\hat{F}$. Next, we apply the map $\us{x}{\ess \ws}\s\ws \leftrightarrow \us{x}{\ws\ws}$ with probability $p$ on those unit cells that belong to subspace (\ref{Eq: Pi}).  We repeat this procedure $t$ times and average over both initial configurations and trajectories (which are now stochastic). An instance of such dynamics is shown in Fig. \ref{Fig: Rule54_snapshot}. While this dynamics can be implemented efficiently via a simple classical Monte Carlo (MC) algorithm, an alternative description can be given using the language of quantum channels.  Quantum channels allow not only for unitary evolution within the system, but \textit{generalized measurements} as well, the latter giving rise to dissipation. The basic ingredient is a set of Kraus operators $\{ \hat{K}_i \}$ that \textit{evolve} a given state described by a density matrix $\hat{\rho}(t) \rightarrow \hat{\rho}(t+1)=\sum_{\mu}\hat{K}_{\mu}\hat{\rho}(t)\hat{K}_\mu^\dagger$ and that satisfy the completeness condition $\sum_\mu \hat{K}_\mu^\dagger \hat{K}_\mu=\hat{\mathbb{1}}$.  A well-known fact is that the choice of these Kraus operators is not unique \cite{nielsen2002quantum}.  The specific choice of Kraus operators for our setup is delegated to \ref{sec_set_Kraus}.  The benefit of a purely quantum mechanical description of the dynamics in terms of Kraus operators is that such time evolution is \textit{exact} and can be in principle simulated via time dependent matrix product operator (tMPO) techniques \cite{schollwock2011density,orus2014practical}. We have ran these simulations finding agreement with MC where possible. In practice we have found that for the time scales involved away from integrability, there is little benefit in using tensor networks.

\subsection{Tracer dynamics} \label{sec_tracer}
\begin{figure}[h!]
\centering
\includegraphics[scale=0.4]{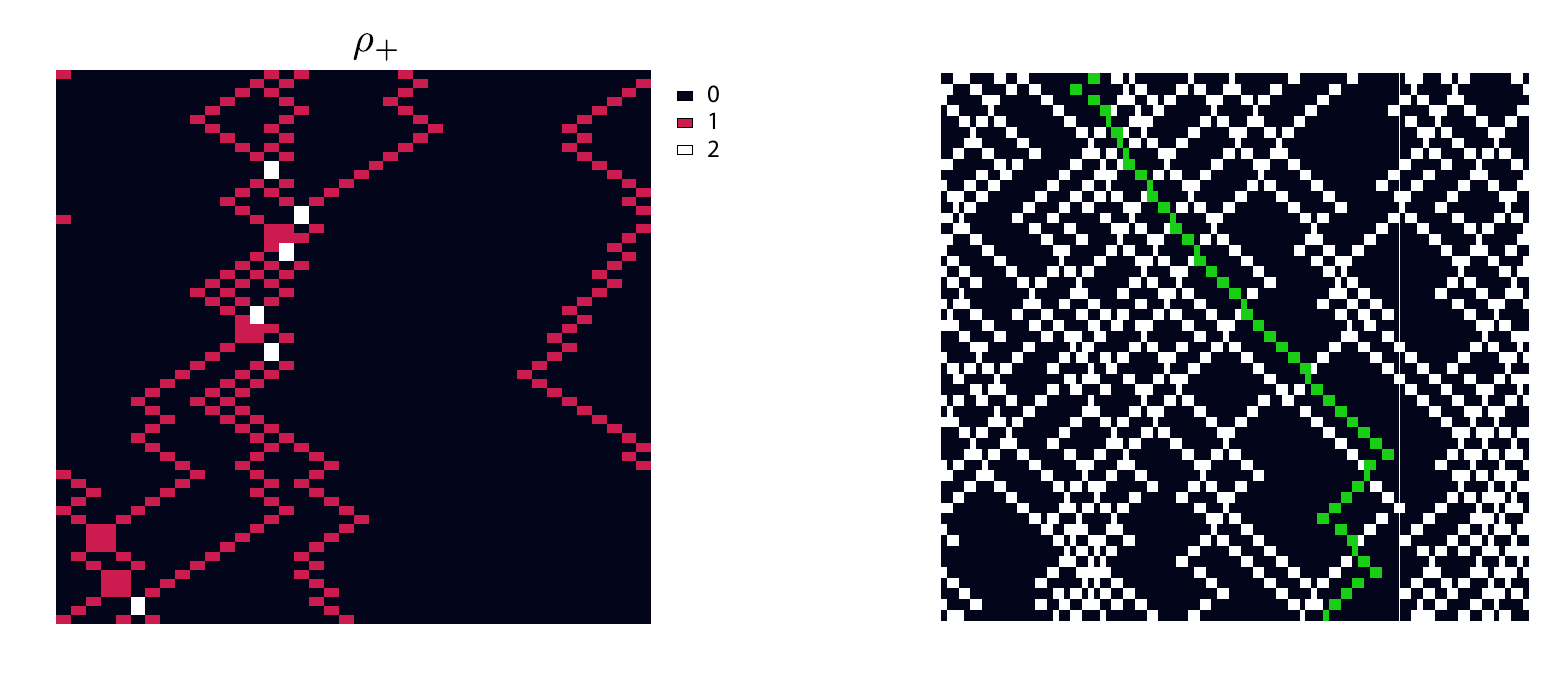}
\caption{\textbf{Snapshot of the noisy Rule 54.}.  Left: Snapshot of density of particles in the presence of noise in Rule 54 at low density $\mu \gg 1$.  Right: dynamics of a tagged particle (in green) near half-filling $\mu=0$.}
\label{Fig: Rule54_snapshot}
\end{figure}

To give some intuition of the transport properties away from integrability we consider the limit $\mu \to \infty$ (low density of particles). Here the dynamics becomes trivial -- interactions are irrelevant and our system is effectively described as a single soliton undergoing a random walk with a mean free path set by the noise, a picture that remains true in general, see left panel of Fig.~\ref{Fig: Rule54_snapshot}.  The imbalance decays with a decay rate given by $\Gamma=2p$ (in the continuous time case) and $\Gamma= -\log(1-2p)$ (in the discrete time case).  The density of particles in turn spreads diffusively. Starting with an initial state $\rho_{+,x}=\delta_{x,0}$ the late time shape for $\rho_+$ is exactly given by the gaussian $\rho_{+,x}=\frac{1}{\sqrt{4\pi \mathcal{D} t}}e^{-\frac{x^2}{4\mathcal{D}t}}$, with a diffusion constant $\mathcal{D}$ given by $\mathcal{D}=1/\Gamma$. 

\begin{figure}[h!]
\centering
\includegraphics[scale=0.27]{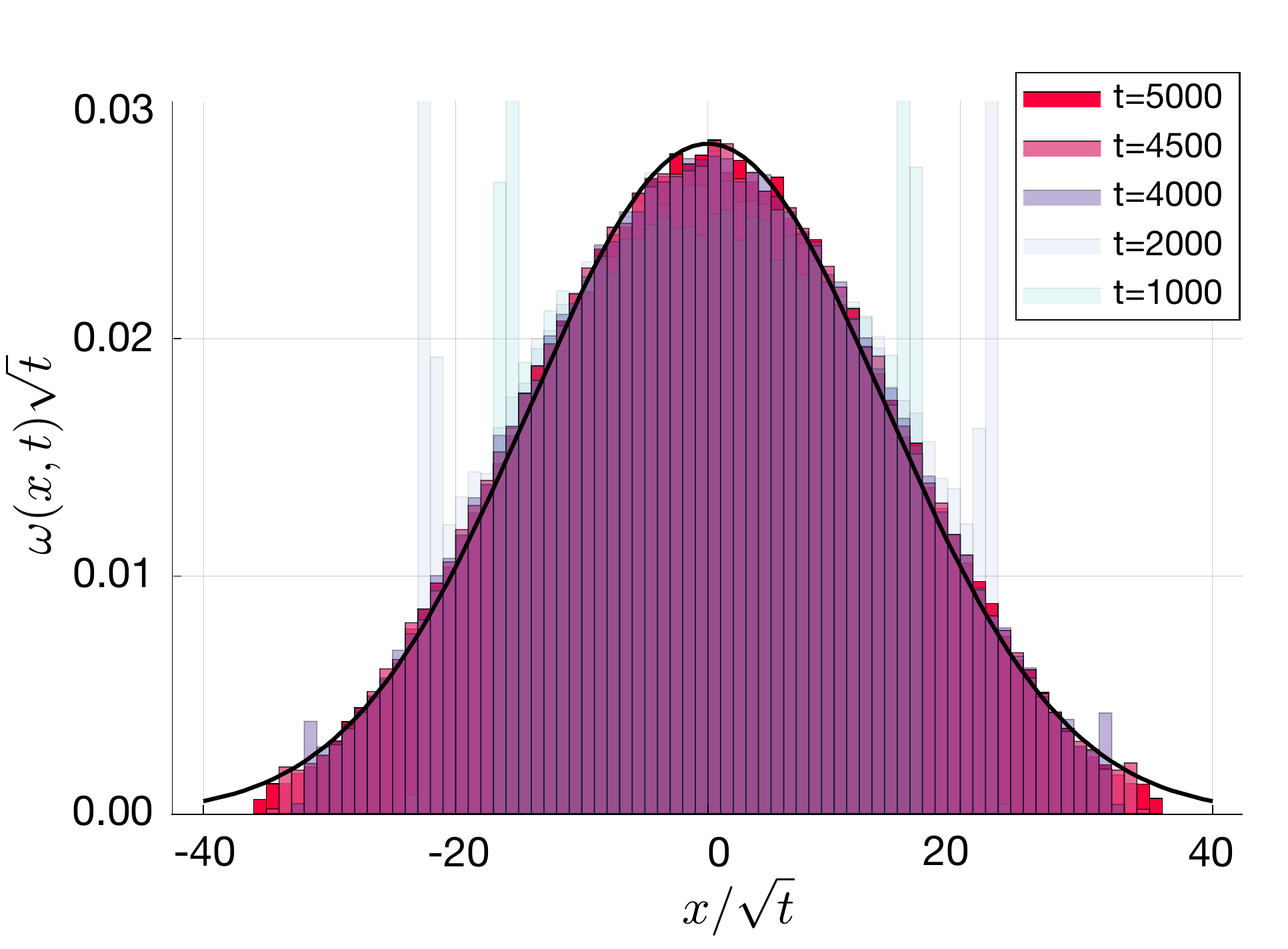}
\caption{\textbf{Tracer distribution profile at half-filling $\mu=0$.}  The distribution follows a normal distribution at long enough times $\sim \mathcal{N}(0,2D^*t)$ (black line) with $D^*$ the self-diffusion constant extracted from (\ref{Eq: self-diffusion}).  Results shown for $p=0.02$ and $\mu=0$ ($D^*=100$). Results for a system size $L=800$ using pbc.}
\label{Fig: Rule54_qp_dist}
\end{figure}
\begin{figure*}[t!]
\centering
\includegraphics[scale=0.21]{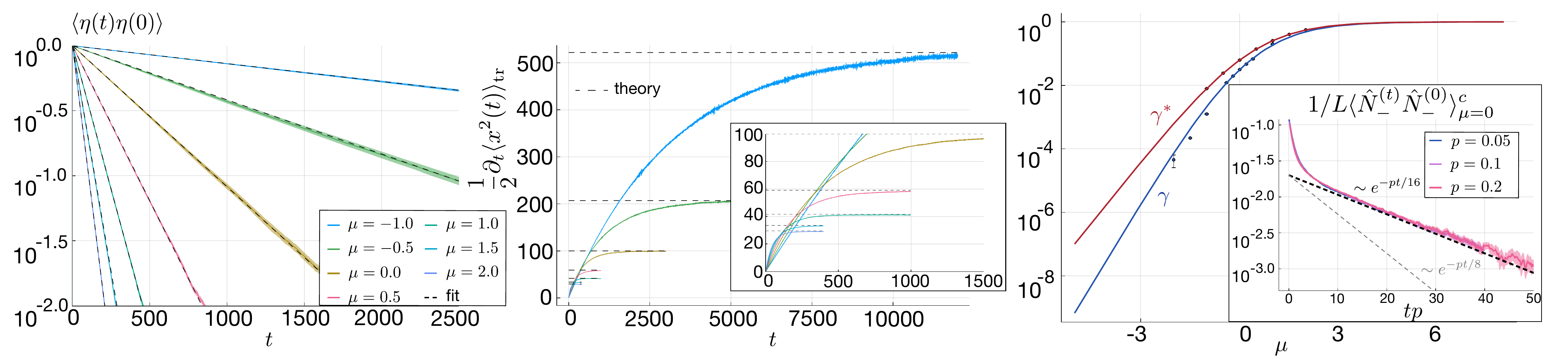}
\caption{\textbf{Tracer vs full dynamics.} Left panel: flavor decay as a function of time for values of $\mu$ that range from $\mu=-1.0$ to $\mu=2.0$ with steps of size $\Delta \mu=0.5$ with $p=0.02$. Black dashed lines correspond to best fits. Middle panel: derivative of tracer variance as a function of time along with theory predictions for the self-diffusion constant $\mathcal{D}^*$ shown in black dashed lines.  Inset shows a zoom over the region of $\mu>0$.  Right panel: flavor decay rate $\gamma^*$ and correlator decay rate (at the longest time scales) $\gamma$ vs $\mu$, including numerical data points extracted from a blind fit.  Inset shows the scaling collapse in the large $t$ limit at $\mu=0$ along with a fit with the conjectured decay rate $\gamma=1/16$.  Also shown for comparision the corresponding flavor decay rate $\gamma^*=1/8$. }
\label{Fig: tracer_vs_full}
\end{figure*}

Away from the trivial limit $\mu \to \infty$ the dynamics consists of a bunch of particles interacting with each other and subject to noise (a \textit{Brownian hard-rod gas} in essence, with fixed bare velocity $\pm 1$).  As a proxy for transport we focus on the dynamics of a tracer.  Studying tracer dynamics is arguably much simpler than \textit{many-body} correlation functions, yet this has proved useful as a first step at determining the nature of transport in integrable systems by exploiting the quasiparticle picture \cite{gopalakrishnan2018hydrodynamics,gopalakrishnan2019kinetic}.  Studying fluctuations of quasiparticle trajectories essentially gives the diagonal components of the diffusion kernel (\ref{eq_diagonal_D}), and as such cannot fully determine the diffusion constant (which requires knowledge of the off-diagonal components as well).  A complete characterization of each tracer $i$ in Rule 54 is given in terms of its \textit{flavor} $\eta_i$ which is a random variable taking the value $+1$ if the tracer is moving right, or $-1$ if it is moving left, and its coordinates $(x,t)$.  A snapshot of what the dynamics looks like for a tagged particle away from integrability is shown on the right panel in Fig. \ref{Fig: Rule54_snapshot}.  We are interested in the dynamics over homogeneous equilibrium background states. The flavor $\eta$ should average to zero, $\langle \eta(t) \rangle =0$ (this is because at equilibrium $\mu_R=\mu_L$ and at any given time,  particles of both flavors are equally likely). As a result we study instead fluctuations $\langle \eta(t)\eta(0)\rangle$. From the theory of Brownian motion we expect this quantity to decay exponentially with a characteristic flavor decay rate $\Gamma^*$ which we seek to determine.  We imagine tagging a right tracer starting at coordinates $(0,0)$. Its ensemble survival probability (the probability for the tracer to turn left) should be proportional to $p \times \langle \hat{\Pi}_{R,x} \rangle$ (this is because the perturbation can only act within the subspace $\hat{\Pi}_{R,x}$).  To fix the proportionality constant we need the constraint that in order for the perturbation to act at $x$, a right particle at $x$ must exist in the first place, hence the survival probability of the tagged right particle should be $\Gamma_R^*=p \times \langle\hat{\Pi}_{R,x}|\hat{\rho}_{R,x}\rangle = p \frac{\langle \hat{\Pi}_{R,x} \rangle}{\langle \hat{\rho}_{R,x} \rangle}=p \frac{\langle \hat{\Pi}_R \rangle }{\langle \hat{\rho}_R \rangle}$, with $\langle\cdot | \cdot \rangle$ denoting conditional expectation value, and where in the last equality we have used the fact that we are considering homogeneous background states,  so that $\langle \hat{\mathcal{O}}_x \rangle = \langle \hat{\mathcal{O}} \rangle$.  An identical argument for the left movers would yield instead $\Gamma_L^*=p \frac{\langle \hat{\Pi}_L \rangle }{\langle \hat{\rho}_L \rangle}$.  At equilibrium $\mu_R=\mu_L =\mu$ the two decay rates are equal and so the flavor decay rate is $\Gamma^*=2p\gamma^*$, with $\gamma^*=\frac{\langle \hat{\Pi}_R \rangle }{\langle \hat{\rho}_R \rangle}=\frac{\langle \hat{\Pi}_L \rangle }{\langle \hat{\rho}_L \rangle}$ and whose expression can be found using the one dimensional transfer matrix of Rule 54 (\ref{Eq: Transfer_matrix})
\begin{equation}
\gamma^*=\frac{e^{3\mu}}{(1+e^{\mu})^2(3+e^{\mu})}.
\end{equation}
We find that our numerical results match perfectly this formula for small enough $p$, as shown in Fig.  \ref{Fig: tracer_vs_full}.  We also note that in the limit of very low filling, this formula reproduces the decay rate expected for a free particle, $\gamma^* \to 1$. With this at hand we can also quantify how tracer particles diffuse.  Each particle will be subject to collisions with other particles and backscattering events that happen at rate $p \gamma^*$. As a result each particle will diffuse with a characteristic self-diffusion constant $\mathcal{D}^*$ (which in general, is different from the \textit{full} diffusion constant $\mathcal{D}$). In other words, the probability distribution $\omega(x,t)$ to find a tracer at coordinates $(x,t)$ assuming it started at the origin, $(0,0)$,  should become a Gaussian that broadens as $\langle x^2(t) \rangle_{\rm tr}\equiv \int x^2 \omega(x,t) \asymp 2\mathcal{D}^*t$, where $\asymp$ denotes the scaling limit $p\to 0^+$, $t \to \infty$, $pt$ fixed.  To determine $\mathcal{D}^*$ we use the following standard formula for a Brownian particle $2\mathcal{D}^*=l^2/\tau^*$ with $l$ the mean-free path and $\tau^*$ the tracer's lifetime. Plugging in $l=v\tau^*$, with $v$ the tracer's velocity whose expression is $v=\frac{(1+e^{\mu})}{(3+e^{\mu})}$ \cite{gopalakrishnan2018hydrodynamics} and $\tau^*=1/p\gamma^*$ we find
\begin{equation} \label{Eq: self-diffusion}
\mathcal{D}^*=\frac{1}{2}v^2\tau^*=\frac{(1+e^{\mu})^4}{e^{3\mu}(3+e^{\mu})}\frac{1}{2p}.
\end{equation} 
The results for the self-diffusion constant are shown in Fig.~\ref{Fig: Rule54_qp_dist}, and Fig.~\ref{Fig: tracer_vs_full} indicating very good agreement with the formula (\ref{Eq: self-diffusion}).

\subsection{Transport in the noisy Rule 54}
Having argued the presence of diffusive dynamics at the level of tracer dynamics, we now briefly discuss transport properties. Using the Kubo formula, we are interested in the two-point function of currents $G_{+,+}(t)=1/L\langle \hat{J}_+^{(t)}\hat{J}_+^{(0)}\rangle^c$.  Recall that in the integrable limit the currents correspond just to the imbalance $\hat{N}_-$, and are thus conserved.  The effects of integrability breaking are already encoded in the temporal evolution of $\hat{N}_-$, which will be the main object of study in what follows [the full current is given locally as $\hat{j}_{+,x+1/2}=(\mathbb{1}-2p\hat{\Pi}_x)\hat{\rho}_{-,x+1/2}$].  To study $G_{+,+}$ we time evolve the total imbalance by one time step, which gives us  
\begin{equation}
\hat{N}_-^{(t+1)}-\hat{N}_-^{(t)}=-2p\hat{\Pi}_-^{(t)},
\end{equation} 
where $\hat{\Pi}_- \equiv \sum_i \hat{\Pi}_{-,i}$ with $\hat{\Pi}_{-,i} \equiv \hat{\Pi}_i \hat{\rho}_{-,i}$.  Already this equation of motion (e.o.m.) tells us that the imbalance should decay. On general grounds,  nonconserved charges are expected to decay exponentially fast with a drift consistent with FGR \cite{PhysRevX.8.021030,mallayya2019prethermalization,friedman2020diffusive,durnin2020non,PhysRevB.103.L060302}.  To extract the drift from here we can evaluate this e.o.m. in an ensemble with slight imbalance, where we take $\mu_R=\mu+\delta$ and $\mu_L=\mu-\delta$ with $\delta \rightarrow 0$.  Alternatively, we can simply study the correlator $\langle \hat{N}_-^{(t)}\hat{N}^{(0)}_- \rangle$ at fixed $\mu_R=\mu_L \equiv \mu$. To lowest order in $p$ we find (see \ref{sec_FGR_appendix} for details on the derivation)
\begin{equation} \label{eq_nmnm_correlator_first_order}
\frac{1}{L}\partial_t \langle \hat{N}_-^{(t)}\hat{N}_-^{(0)} \rangle =-4p \frac{e^{3\mu}}{(1+e^\mu)^3(3+e^\mu)} + \mathcal{O}(p^2),
\end{equation}
where the time derivative is taken to be discrete. Eq. (\ref{eq_nmnm_correlator_first_order}) is exact up to leading order in $p$, and thus is analagous to FGR (\ref{eq_FGR}). However, for our noisy Rule 54 model the subleading terms $\mathcal{O}(p^2)$ cannot be disregarded, as shown in \ref{sec_2nd_order_corrections} thereby indicating a breakdown of perturbation theory in the noisy FFA, and ultimately implying a breakdown of FGR in our setup. More precisely, the terms $\mathcal{O}(p^2)$ involve sums over time of correlators which approach a constant at long times. The presence of such a ``Drude weight'' in those correlators imply that the higher-order terms actually scale with time, and cannot be neglected at long times even if $p \ll 1$. 
A possible way to interpret this result is that Rule 54 fails to equilibrate on its own.  The lack of a dephasing mechanism in the model is what prevents from truncating the series expansion to leading order in the perturbation coupling $p$.  In \ref{sec_2nd_order_corrections} we indeed show by computing explicitly the second order corrections (as well as the third order corrections) that they grow in time as $\mathcal{O}(t^2)$ ($\mathcal{O}(t^3)$), thereby invalidating any perturbative analysis in the vein of FGR.  One can show however that a proper resummation of the perturbative series reveals the following scaling form of the two-point function for the currents (see \ref{sec_2nd_order_corrections})
\begin{equation}
G_{+,+}(t) \asymp F(pt),
\end{equation}
with $F$ a function characterizing transport that would be interesting to characterize in future work. While the function $F$ decays exponentially for $pt \gg1$, we emphasize that the diffusion constant depends on the whole scaling function $F$, and not on a single decay rate. Properly characterizing this function would likely involve understanding the decay of all the other conserved quantities of Rule 54, even if those decouple from its hydrodynamics in the integrable case.

\section{Conclusion} \label{sec_conclusion}
We have studied a version of the Rule 54 that breaks the integrability of the model. This leads to diffusion of the conserved charges and decay of the nonconserved charges.  The particular simplicity of the model allowed us to study in detail tracer dynamics in this system. Transport on the other hand is nontrivial as a result of the rather slow dynamics present in this setup. Currents decay through a set of decay rates which cannot be extracted by means of naive perturbation theory, and this prevents us from having access to analytical formulae for the various transport coefficients.  It would be interesting to understand how generic the present analysis is in the context of other systems under constrained dynamics and whether a more clear cut link between the dynamics of the tracer and transport can be made. 
\ack
This work was supported by the National Science Foundation under NSF Grant No. DMR-1653271 (S.G.), the US Department of Energy, Office of Science, Basic Energy Sciences, under Early Career Award No. DE-SC0019168 (R.V. and J.L.), and the Alfred P. Sloan Foundation through a Sloan Research Fellowship (R.V.).
\appendix
\section{Quantum channel description of noisy Rule 54}
\subsection{Setting up the Kraus operators} \label{sec_set_Kraus}
Our model is defined as follows: after each unitary step $\hat{F}$, we convert with certain probability $p$ a right mover into a left mover and viceversa. This will preserve the total number of particles $N_+$, but not the imbalance $N_-$. It is easy to realize that the minimum set of Kraus operators implementing this operation at each unit cell $j$ is two and must be of the form 
\numparts
\begin{eqnarray}
\label{Eq: K_1}
\hat{K}_{\mu_j=1}=\mathbb{1}_j-\hat{\Pi}_j+\sqrt{1-p}\hat{\Pi}_j, \\
\hat{K}_{\mu_j=2}=\sqrt{p}\hat{\Pi}_j\hat{S}_j\hat{\Pi}_j,
\end{eqnarray}
\endnumparts
with $j \in \mathbb{Z}$. Here $\hat{\Pi}$ projects onto a given (to be determined) subspace composed of right/left movers while $\hat{S}$ swaps a right mover and a left mover within said subspace. The identity element in (\ref{Eq: K_1}) guarantees the completeness condition of Kraus operators. A full circuit layer is thus given by $\hat{\rho}(t+1)=\sum_{\{\vec{\mu} \}}\hat{K}_{\vec{\mu}} \hat{F}^\dagger\hat{\rho}(t)\hat{F}\hat{K}^\dagger_{\vec{\mu}}$, with $\vec{\mu}=(\mu_1,\mu_2,...,\mu_L)$ and we define $\hat{K}_{\vec{\mu}}\equiv \bigotimes_{i=1}^L \hat{K}_{\mu_i}$. The structure of the model, in particular the fact that the density of right and left movers depend on three sites (two unit cells) already tells us that we should make sure that not only at a given unit cell $N_+$ is preserved, but also at its adjacent cells.  To keep matters simple we consider the operator $\hat{S}_j$ that swaps the states $|\us{$j$}{\ws \ws}\s\bss \rangle$ (corresponding to one right mover) and $|\us{$j$}{\bss \ws}\s\ws\rangle$ (corresponding to one left mover). That is, $\hat{S}_j=\hat{\sigma}_j^+\us{j+1/2}{\ws}\hat{\sigma}_{j+1}^- + \rm h.c.$.  Having found $\hat{S}$ we can determine $\hat{\Pi}$. After some trial and error we find that the most general projector that still preserves $N_+$ must be of the form 
\begin{eqnarray} \label{Eq: projector}
\hat{\Pi}_i = \alpha (\us{$i-1$}{\ess\bss}\s\us{$i$}{\ws\ws}\s\us{$i+1$}{\bss\bss}+ \us{$i-1$}{\ess\bss}\s\us{$i$}{\bss\ws}\s\us{$i+1$}{\ws\bss}) \nonumber\\
+\beta(\us{$i-1$}{\ws\ws}\s\us{$i$}{\ws\ws}\s\us{$i+1$}{\bss\ws}\s\us{$i+2$}{\ws\ess}+\us{$i-1$}{\ws\ws}\s\us{$i$}{\bss\ws}\s\us{$i+1$}{\ws\ws}\s\us{$i+2$}{\ws\ess})\\ 
+\gamma (\us{$i-1$}{\bss\ws}\s\us{$i$}{\ws\ws}\s\us{$i+1$}{\bss\ws}\s\us{$i+2$}{\bss\ess}+\us{$i-1$}{\bss\ws}\s\us{$i$}{\bss\ws}\s\us{$i+1$}{\ws\ws}\s\us{$i+2$}{\bss\ess}) \nonumber
\end{eqnarray}
with $\alpha,\beta,\gamma \in \{0,1\}$. Note that these Kraus operators are symmetric, $\hat{K}^\dagger_{\mu_i}=\hat{K}^T_{\mu_i}=\hat{K}_{\mu_i}$. This work considers the simplest case with $\alpha=1,\beta=\gamma=0$, for which Kraus operators mutually commute with each other, $[\hat{K}_{\mu_i}, \hat{K}_{\mu_j}]=0$ for any $i,j$ (this is not true for the other possible choices). This will permit us to \textit{encode} dissipation in a single circuit layer. Note that this circuit has a natural interpretation as an extension of the Rule 54 cellular automaton when including noise of strength $p$ and as such we can simulate it classically as well.  For details on the numerical implementation of the quantum channel, we refer to ``Numerical details" section of the Appendix.
\subsection{Perturbative expansion of the quantum channel} \label{sec_FGR_appendix}
Let $\vec{\rho}_-$ the density of left movers in vector form, i.e. $\vec{\rho}_-\equiv 1/L\vec{N}_-$.  Time evolution for $t$ time steps gives $\vec{\rho}_-^{(t)}=\Phi^{(t)}\vec{\rho}_-^{(0)}$ with the map 
\begin{equation}
\Phi^{(t)} \equiv \left(\sum_{\{\vec{\mu}\}} \hat{K}_{\vec{\mu}}\hat{F}^\dagger\otimes \hat{K}_{\vec{\mu}}\hat{F}^\dagger\right)^t,
\end{equation}
where we have implicitly made use of the properties $\hat{K}_{\vec{\mu}}^\dagger=\hat{K}_{\vec{\mu}}^T=\hat{K}_{\vec{\mu}}$ and $\hat{F}^\dagger=\hat{F}^T$. At this step it is useful to start a perturbative expansion in $p$ of $\sum_{\{\vec{\mu}\}}\hat{K}_{\vec{\mu}}\otimes \hat{K}_{\vec{\mu}}$.  Let us denote $\hat{\Theta}_i\equiv \hat{\Pi}_i \hat{S}_i=\hat{\Pi}_i\hat{S}_i\hat{\Pi}_i$.  Carrying out the expansion we get
\begin{equation}
\fl \sum_{\{\vec{\mu}\}}\hat{K}_{\vec{\mu}}\otimes \hat{K}_{\vec{\mu}}=\hat{K}_{(0,0,...,0)}\otimes \hat{K}_{(0,0,...,0)}+\sum_{i=1}^L\hat{K}_{(0,0,...,0,\underset{$i$}{\scriptsize{1}},0,...,0)}\otimes \hat{K}_{(0,0,...,0,\underset{$i$}{\scriptsize{1}},0,...,0)} +\mathcal{O}(p^2).
\end{equation} 
Let us consider the first term. This gives us
\begin{eqnarray}
\fl \hat{K}_{(0,0,...,0)}\otimes \hat{K}_{(0,0,...,0)}=\prod_{i=1}^L\left(\mathbb{1}+\hat{\Pi}_i\sum_{n=1}^\infty (-1)^np^n\ {\frac{1}{2} \choose n}\right)\otimes \prod_{i=1}^L\left(\mathbb{1}+\hat{\Pi}_i\sum_{n=1}^\infty (-1)^np^n\ {\frac{1}{2} \choose n}\right) \nonumber\\
=\mathbb{1}-\frac{p}{2}\sum_{i=1}^L\left(\mathbb{1}\otimes \hat{\Pi}_i+\hat{\Pi}_i\otimes \mathbb{1}\right) +\mathcal{O}(p^2). 
\end{eqnarray}
The second term takes also a very simple form 
\begin{eqnarray} \label{Eq: 2nd_term}
\hat{K}_{(0,0,...,0,\underset{$i$}{\scriptsize{1}},0,...,0)}\otimes \hat{K}_{(0,0,...,0,\underset{$i$}{\scriptsize{1}},0,...,0)}=p\hat{\Theta}_i\otimes\hat{\Theta}_i  + \mathcal{O}(p^{2}).
\end{eqnarray}
Thus to first order in $p$ we have
\begin{equation} \label{Eq: Kraus_ops_expansion}
\sum_{\{\vec{\mu}\}}\hat{K}_{\vec{\mu}}\otimes \hat{K}_{\vec{\mu}}=\mathbb{1}+p\sum_{i=1}^L\left(\hat{\Theta}_i\otimes\hat{\Theta}_i  -\frac{1}{2}\left(\mathbb{1}\otimes \hat{\Pi}_i+\hat{\Pi}_i\otimes \mathbb{1}\right)\right)+\mathcal{O}(p^2).
\end{equation}
Plugging this into $\Phi^{(t)}$ we get
\begin{eqnarray}
\fl \Phi^{(t)}=(\hat{F}^\dagger \otimes \hat{F}^\dagger)^t+ \nonumber \\
\fl +p\sum_{i=1}^L\sum_{n=1}^t (\hat{F}^\dagger \otimes \hat{F}^\dagger)^{t-n}\left(\hat{\Theta}_i\otimes\hat{\Theta}_i  -\frac{1}{2}\left(\mathbb{1}\otimes \hat{\Pi}_i+\hat{\Pi}_i\otimes \mathbb{1}\right)\right)(\hat{F}^\dagger \otimes \hat{F}^\dagger)^n +\mathcal{O}(p^2).
\end{eqnarray}
Applying this to the local density of imbalance which we take it to be $\hat{\rho}_{-,l}$, with $l \in [1,L]$ we have
\begin{equation} \label{Eq: rho_-}
\fl \hat{\rho}_{-,l}^{(t)}=\hat{\rho}_{-,l}^{0(t)}+p\sum_{i=1}^L\sum_{n=1}^t (\hat{F}^{\dagger})^{t-n}\hat{\Theta}_i \hat{\rho}^{0(n)}_{-,l}\hat{\Theta}_i \hat{F}^{t-n}-(\hat{F}^\dagger)^{t-n}\hat{\Pi}_i \hat{\rho}^{0(n)}_{-,l}\hat{\Pi}_i\hat{F}^{t-n} +\mathcal{O}(p^2),
\end{equation}
where $\hat{\rho}_{-,l}^{0(n)}\equiv (\hat{F}^{\dagger})^n\hat{\rho}_{-,l}^{(0)}\hat{F}^n$ and we have made use of $[\hat{\rho}_{-,l}^{0(n)},\hat{\Pi}_i]=0$ $\forall i,l,n$.   We seek an e.o.m. for the decaying charge. Let $\hat{N}_-=\sum_l \hat{\rho}_{-,l}$. Its discrete time derivative reads 
\begin{equation} \label{Eq: 1st_order_p_Nminus}
\fl \hat{N}_-^{(t+1)}-\hat{N}_-^{(t)}=p\sum_{i=1}^L (\hat{F}^\dagger)^t \hat{\Theta}_i \hat{N}_-^{(0)} \hat{\Theta}_i \hat{F}^t-(\hat{F}^\dagger)^t \hat{\Pi}_i \hat{N}_-^{(0)}\hat{\Pi}_i \hat{F}^t + \mathcal{O}(p^2),
\end{equation}
where have used conservation of the global charge under the unperturbed dynamics, $[\hat{F},\hat{N}_-]=0$. Multiplying both sides of (\ref{Eq: 1st_order_p_Nminus}) by $\hat{N}_-^{(0)}$ and averaging over an homogeneous Gibbs state of fixed $\mu$, $\langle \cdots \rangle \equiv 1/Z \tr[\cdots e^{-\mu \hat{N}_+}]$, we get
\begin{eqnarray}
\fl \langle(\hat{N}_-^{(t+1)}-\hat{N}_-^{(t)})\hat{N}_-^{(0)}\rangle = p\sum_{i=1}^L\langle\hat{\Theta}_i\hat{N}_-^{(0)}\hat{\Theta}_i\hat{N}_-^{(0)}\rangle-\langle\hat{\Pi}_i\hat{N}_-^{(0)}\hat{\Pi}_i\hat{N}_-^{(0)}\rangle+\mathcal{O}(p^2) \nonumber\\
=-2p\langle \hat{\Pi}_-\hat{N}_-^{(0)}\rangle +\mathcal{O}(p^2)\\
=-2p\langle \hat{\Pi}\rangle+\mathcal{O}(p^2),  \nonumber
\end{eqnarray}
where in the first line we have made use of $[\hat{F},\hat{N}_+]=0$, in the second line we have used 
\begin{equation} \label{Eq: SWAP}
\hat{\Theta}_i\hat{\rho}_{-,j}\hat{\Theta}_i=(1-2\delta_{i,j})\hat{\rho}_{-,j}\hat{\Pi}_i
\end{equation} 
and defined $\hat{\Pi}_-=\sum_i \hat{\Pi}_{-,i}$ with $\hat{\Pi}_{-,i}\equiv \hat{\Pi}_i \hat{\rho}_{-,i}$, and in the last line we have made use of $\hat{\rho}_{-,i}\hat{\Pi}_{-,i}=\hat{\Pi}_i$, and $\hat{\Pi}= \sum_i \hat{\Pi}_i$.  Using the transfer matrix (\ref{Eq: Transfer_matrix}) we get 
\begin{equation}
\langle \hat{\Pi}_i \rangle=2\frac{e^{3\mu}}{(1+e^{\mu})^3(3+e^{\mu})}, 
\end{equation}
and so the derivative of the correlator at order $p$ reads (taking time continuous)
\begin{equation} \label{Eq: N_-_derivative_1st_order}
\partial_t \frac{1}{L}\langle \hat{N}_-^{(t)} \hat{N}_-^{(0)} \rangle = -4p\frac{e^{3\mu}}{(1+e^{\mu})^3(3+e^{\mu})} +\mathcal{O}(p^2).
\end{equation}
In the following we shall see that the next to leading order terms $\mathcal{O}(p^2)$ cannot in fact be disregarded, as the small parameter controlling this expansion is $pt$ and not $p \tau^*$: the r.h.s.  \textit{does} depend on time through the term $\mathcal{O}(p^2)$.  We shall discuss these higher order corrections next but before, let us state that in (\ref{Eq: N_-_derivative_1st_order}) at $t=0$, terms of $p^2$ and higher order are zero (in general, at any time step $t$, only terms of at most order $p^{t}$ can contribute; see below) and so one can extract a decay rate as 
\begin{equation}
\partial_t \langle \hat{N}_-^{(t)}\hat{N}_-^{(0)}\rangle|_{t=0} = -\Gamma_{\rm s.t.} \langle \hat{N}_-^{(t)}\hat{N}_-^{(0)}\rangle|_{t=0},
\end{equation} 
where $\Gamma_{\rm s.t.}=2p\gamma_{\rm s.t.}$ with 
\begin{equation}
\gamma_{\rm s.t.}=\frac{\langle \hat{\Pi} \rangle}{LG_{+,+}}=\frac{e^{2\mu}(3+e^{\mu})}{(1+e^{\mu})^3}.
\end{equation}
where $G_{+,+}=1/L\langle \hat{N}_-^2 \rangle$ is the current-current correlator of number of solitons in the integrable limit. The subscript s.t.~here is to denote \textit{short time}, given that the extracted decay rate is only an approximation that is valid at very short times (strictly, at $t=0$). 

%Before jumping onto the study of higher order corrections let us remind that under general conditions, either studying the evolution of a charge $\hat{\mathcal{Q}}_\beta$ within linear response, $\langle \hat{\mathcal{Q}}_\beta^{(t)} \rangle_{\rm LR}$, or its fluctuations $\langle \hat{\mathcal{Q}}_\beta^{(t)}\hat{\mathcal{Q}}_\beta^{(0)}\rangle$, provide the same information. More precisely, assume $\langle \hat{\mathcal{Q}}_\beta^{(t)}\rangle=0$. Let $\langle \cdot \rangle_{\rm LR} \equiv \frac{1}{\tr[e^{-\sum_{\alpha \neq \beta} \mu_\alpha \hat{\mathcal{Q}}_\alpha - \mu_\beta \hat{\mathcal{Q}}_\beta}]}\tr[\cdot e^{-\sum_{\alpha \neq \beta} \mu_\alpha \hat{\mathcal{Q}}_\alpha - \mu_\beta \hat{\mathcal{Q}}_\beta}]$ with $|\mu_\beta| \ll 1$, then $\partial_{\mu_\beta} \langle \hat{\mathcal{Q}}_{\beta}^{(t)}\rangle_{\rm LR}\big|_{\mu_\beta=0}=\langle \hat{\mathcal{Q}}_\beta^{(t)}\hat{\mathcal{Q}}_\beta^{(0)}\rangle$. Thus, in what follows we will use interchangeably e.v.s of $\hat{N}_-$ evaluated within linear response or study its fluctuations. 

\section{Perturbation theory beyond first order} \label{sec_2nd_order_corrections}
\subsection{2nd order corrections}
Let us take a closer look now at the corrections of order $\mathcal{O}(p^2)$. We have 
\begin{eqnarray} \label{Eq: expansion}
\fl \sum_{\{\vec{\mu}\}}\hat{K}_{\vec{\mu}}\otimes \hat{K}_{\vec{\mu}}=\hat{K}_{(0,0,...,0)}\otimes \hat{K}_{(0,0,...,0)}+\sum_{i=1}^L\hat{K}_{(0,0,...,0,\underset{i}{\scriptsize{1}},0,...,0)}\otimes \hat{K}_{(0,0,...,0,\underset{i}{\scriptsize{1}},0,...,0)} + \nonumber \\
+\sum_{i=1}^{L-1}\sum_{j>i}^L\hat{K}_{(0,0,...,0,\underset{i}{\scriptsize{1}},0,...,0,\underset{j}{\scriptsize{1}},0,...,0)}\otimes \hat{K}_{(0,0,...,0,\underset{i}{\scriptsize{1}},0,...,0,\underset{j}{\scriptsize{1}},0,...,0)} + \mathcal{O}(p^3).
\end{eqnarray} 
The first term in the r.h.s. gives us 
\begin{eqnarray}
\fl \hat{K}_{(0,0,...,0)}\otimes \hat{K}_{(0,0,...,0)}=\prod_{i=1}^L\left(\mathbb{1}-\frac{p}{2}\hat{\Pi}_i-\frac{p^2}{8}\hat{\Pi}_i\right)\otimes \prod_{i=1}^L\left(\mathbb{1}-\frac{p}{2}\hat{\Pi}_i-\frac{p^2}{8}\hat{\Pi}_i\right) +\mathcal{O}(p^3) \nonumber \\
=\mathbb{1}-\frac{p}{2}\sum_{i=1}^L(\mathbb{1}\otimes \hat{\Pi}_i+\hat{\Pi}_i\otimes \mathbb{1})+\\
+\frac{p^2}{4}\sum_{i=1}^L \left(\hat{\Pi}_i\otimes \sum_{j=1}^L\hat{\Pi}_j-\left(\frac{\hat{\Pi}_i}{2}-\hat{\Pi}_i\sum_{j>i}^L\hat{\Pi}_j\right)\otimes \mathbb{1}-\mathbb{1}\otimes \left(\frac{\hat{\Pi}_i}{2}-\hat{\Pi}_i\sum_{j>i}^L\hat{\Pi}_j\right) \right) \nonumber \\
+\mathcal{O}(p^3). \nonumber
\end{eqnarray}
We also have, using (\ref{Eq: 2nd_term}):
\begin{equation}
\fl \hat{K}_{(0,0,...,0,\underset{i}{\scriptsize{1}},0,...,0)} \otimes \hat{K}_{(0,0,...,0,\underset{i}{\scriptsize{1}},0,...,0)}=p\hat{\Theta}_i\otimes \hat{\Theta}_i-\frac{p^2}{2}\left(\hat{\Theta}_i\otimes \hat{\Theta}_i\right)\sum_{j\neq i}^L\left(\hat{\Pi}_j\otimes \mathbb{1}+\mathbb{1}\otimes \hat{\Pi}_j\right)+\mathcal{O}(p^3).
\end{equation}
For the last term in (\ref{Eq: expansion}) we have:
\begin{equation}
\hat{K}_{(0,0,...,0,\underset{i}{\scriptsize{1}},0,...,0,\underset{j}{\scriptsize{1}},0,...,0)}\otimes \hat{K}_{(0,0,...,0,\underset{i}{\scriptsize{1}},0,...,0,\underset{j}{\scriptsize{1}},0,...,0)} =p^2\hat{\Theta}_i\hat{\Theta}_j \otimes \hat{\Theta}_i\hat{\Theta}_j+\mathcal{O}(p^3).
\end{equation}
Let us now rewrite $\Phi^{(t)}$ as follows 
\begin{eqnarray} \label{Eq: W_expansion}
\fl \Phi^{(t)}=((\mathbb{1}+pB+p^2C)A)^t \nonumber\\
\fl =A^t+p\sum_{n=1}^tA^{t-n}BA^{n}+p^2\left(\sum_{n=1}^t A^{t-n}CA^{n}+\sum_{n=1}^{t-1}\sum_{m>n}^tA^{t-m}BA^{n}BA^{m-n}\right)+\mathcal{O}(p^3),
\end{eqnarray}
where we define $A\equiv \hat{F}^\dagger \otimes \hat{F}^\dagger$, $B\equiv\sum_{i=1}^L B_{0,i}+B_{1,i}$, and $C\equiv \sum_{i=1}^L C_{0,i}+C_{1,i}+C_{2,i}$ and
\begin{eqnarray} \label{Eq: B_C}
\fl B_{0,i}\equiv -\frac{1}{2}\left(\mathbb{1}\otimes \hat{\Pi}_i+\hat{\Pi}_i\otimes \mathbb{1}\right), \nonumber \\
\fl B_{1,i}\equiv \hat{\Theta}_i \otimes \hat{\Theta}_i,\nonumber \\
\fl C_{0,i} \equiv \frac{1}{4}\left(\hat{\Pi}_i\otimes \sum_{j=1}^L\hat{\Pi}_j-\left(\frac{\hat{\Pi}_i}{2}-\hat{\Pi}_i\sum_{j>i}^L\hat{\Pi}_j\right)\otimes \mathbb{1}-\mathbb{1}\otimes \left(\frac{\hat{\Pi}_i}{2}-\hat{\Pi}_i\sum_{j>i}^L\hat{\Pi}_j\right)\right),\\
\fl C_{1,i}\equiv -\frac{1}{2}\left(\hat{\Theta}_i\otimes \hat{\Theta}_i\right)\sum_{j\neq i}^L\left(\hat{\Pi}_j\otimes \mathbb{1}+\mathbb{1}\otimes \hat{\Pi}_j\right), \nonumber \\
\fl C_{2,i}\equiv \sum_{j>i}^L \hat{\Theta}_i\hat{\Theta}_j \otimes \hat{\Theta}_i\hat{\Theta}_j.\nonumber
\end{eqnarray}

\begin{figure}
\centering
\includegraphics[scale=0.5]{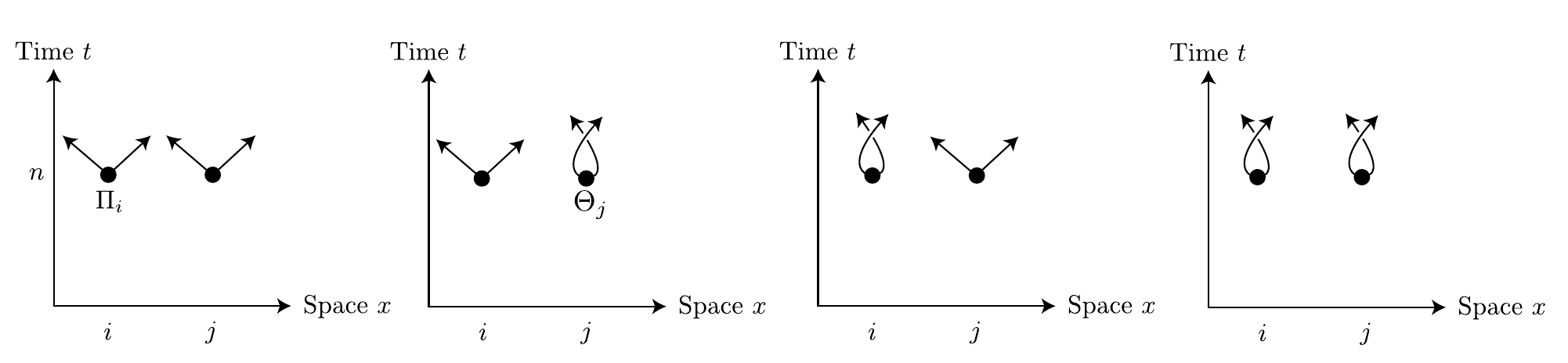}
\caption{Pictorial description of each of the terms in Eqs.  (\ref{Eq: C_term_1}, \ref{Eq: C_term_2}, \ref{Eq: C_term_3}). Leftmost figure projects the density of imbalance $\hat{\rho}_{-,l}$ onto the subspace $\hat{\Pi}$ at locations $i$ and $j$ (Eq.  (\ref{Eq: C_term_1})). The two middle figures consider projecting onto one such subspace and swapping right with left movers at only one location (Eq. (\ref{Eq: C_term_2})). The rightmost figure swaps right with left movers at both locations $i$ and $j$ (Eq. (\ref{Eq: C_term_3})).}
\label{Fig: pic1}
\end{figure}
\begin{figure}
\centering
\includegraphics[scale=0.5]{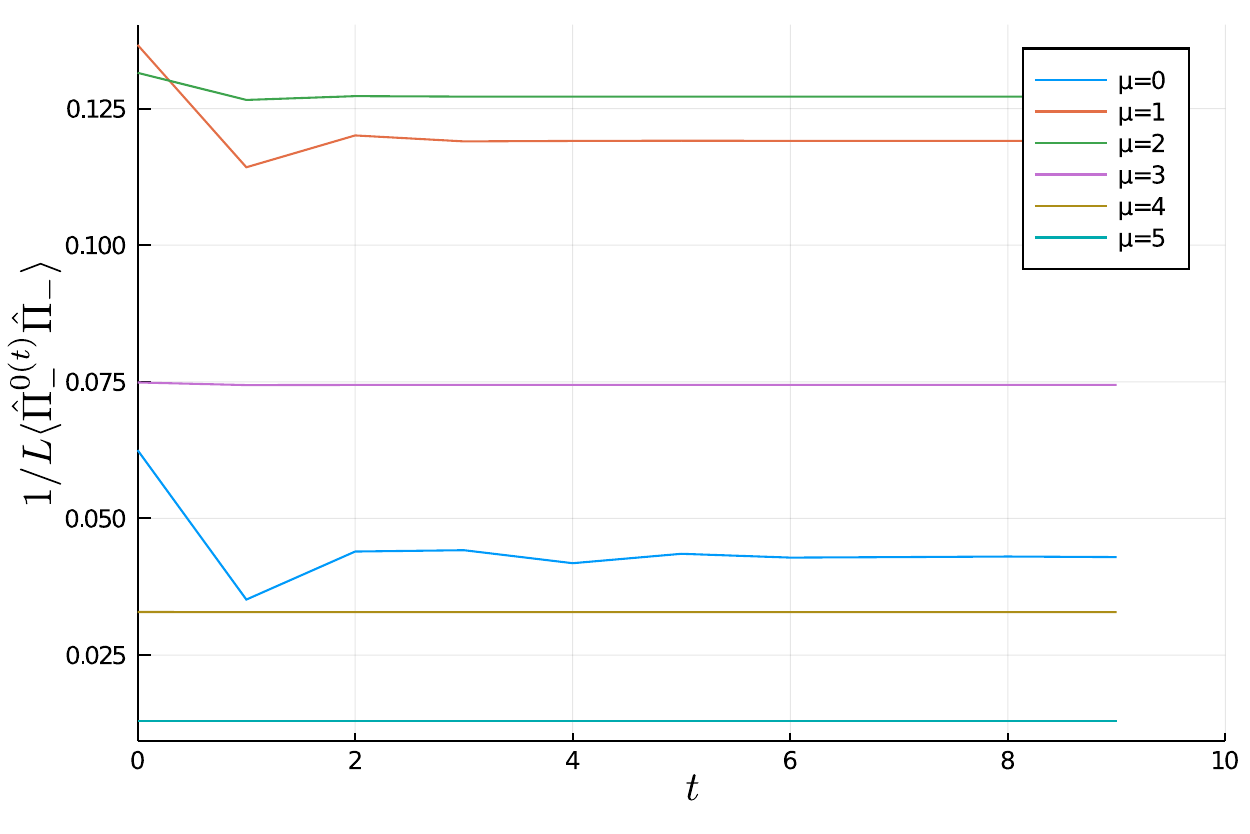}
\caption{2-point function $\frac{1}{L}\langle \hat{\Pi}_-^{0(t)}\hat{\Pi}_- \rangle$.}
\label{Fig: Pi_corr}
\end{figure}
\begin{figure}
\includegraphics[scale=0.25]{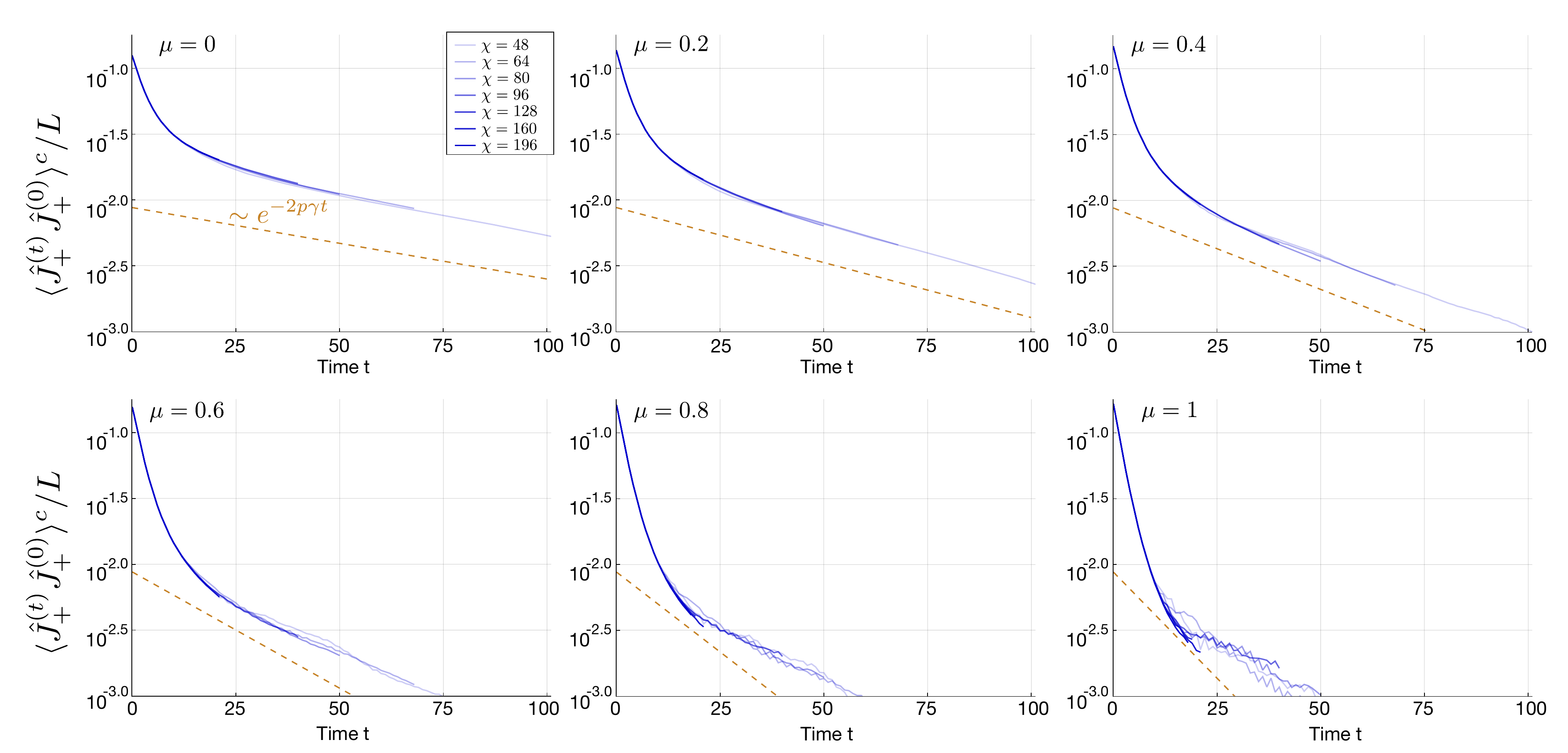}
\caption{\textbf{Decay of currents in the noisy Rule 54 CA from tMPO numerics.} Results for various values of $\mu$ for noise strength $p=0.2$ and theoretical asymptotic decay with $\Gamma=2p\gamma$, with $\gamma=(1-n)\gamma^*$.  As $\mu$ increases the tMPO simulations become more unstable (require higher bond dimensions to achieve convergence).}  
\label{fig_JpJp_MPO}
\end{figure}

At this point it is best to apply $\Phi^{(t)}$ from (\ref{Eq: W_expansion}) to the local imbalance density, $\hat{\rho}_{-,l}^{(0)}$, and sum over the space coordinate at the end.  Focusing now on the $\mathcal{O}(p^2)$ terms, consider first the term $\sum_{i=1}^LA^{t-n}C_i A^{n}$.  A priori every single term in $C_i$ contributes when taking the expectation value over an homogeneous state.  Using this already yields (with the convention $\vec{\rho} \cong \hat{\rho}$) 
\numparts
\begin{eqnarray}\label{Eq: C_term_1} 
(A^{t-n}C_{0,i}A^{n}) \vec{\rho}_{-,l}^{(0)} \cong \sum_{j>i}^L(\hat{F}^\dagger)^{t-n}\hat{\Pi}_i\hat{\Pi}_j\hat{\rho}_{-,l}^{0(n)} \hat{\Pi}_j \hat{\Pi}_i \hat{F}^{t-n},\\
\label{Eq: C_term_2} 
(A^{t-n}C_{1,i}A^{n})\vec{\rho}_{-,l}^{(0)}  \cong -\sum_{j\neq i}^L (\hat{F}^\dagger)^{t-n}\hat{\Theta}_i \hat{\Pi}_j \hat{\rho}_{-,l}^{0(n)}\hat{\Pi}_j\hat{\Theta}_i \hat{F}^{t-n},\\
\label{Eq: C_term_3} 
(A^{t-n}C_{2,i}A^{n})\vec{\rho}_{-,l}^{(0)} \cong \sum_{j>i}^L (\hat{F}^\dagger)^{t-n}\hat{\Theta}_i\hat{\Theta}_j \hat{\rho}_{-,l}^{0(n)}\hat{\Theta}_j\hat{\Theta}_i \hat{F}^{t-n}, 
\end{eqnarray}
\endnumparts
where as before we have made use of $[\hat{\rho}_{-,l}^{0(n)},\hat{\Pi}_i]=0$.  This equation has a simple interpretation: at order $\mathcal{O}(p^2)$ we have three contributions \underline{at any given} time all of which arise after \textit{freezing} two mutually disjoint regions on which the projector $\hat{\Pi}$ acts, that is,  two mutually disjoint regions consisting each of them with a single right mover or a left mover. In the first scenario (top most relation) the total imbalance is conserved within each of these subspaces.  The second case reflects instead the swapping of one right mover with a left mover at one of such locations, leading to the net change of two in the imbalance. The last expression indicates swapping right and left quasiparticles in these two regions at any given time, leading to a net change of four in the imbalance. These different processes are shown pictorially in Fig. \ref{Fig: pic1}.  We can now simplify matters by summing over the space coordinate for the imbalance density, in order to make use of conservation of the global charge $N_{-}$ under the unperturbed dynamics.  It is easy to see that (\ref{Eq: C_term_1})+(\ref{Eq: C_term_2})+(\ref{Eq: C_term_3})=0, $(A^{t-n}CA^n)\vec{N}_-=0$, when using (\ref{Eq: SWAP}). %Using as well translation invariance and relation (\ref{Eq: SWAP})
%we arrive at 
%\begin{eq} \label{Eq: ACA}
%\begin{split}
%\langle ((A^{t-n}CA^{n})) \vec{N_{-}}^{(0)}\rangle_{LR} &\cong t \sum_{i=1}^L\sum_{j<i}^L \left(\sum_{l=1}^L \langle \hat{\Pi}_i \hat{\Pi}_j \hat{\rho}_{-,l}\hat{\Pi}_j \hat{\Pi}_i \rangle_{LR} -\sum_{l=1}^L \langle \hat{\Theta}_i \hat{\Pi}_j \hat{\rho}_{-,l} \hat{\Pi}_j \hat{\Theta}_i \rangle_{LR} \right.\\
%& -2 \langle \hat{\Theta}_j \hat{\Pi}_i \hat{\rho}_{-,i}\hat{\Pi}_i \hat{\Theta}_j \rangle_{LR} \bigg ).
%\end{split}
%\end{eq}
%After some tedious but straightforward algebra (using identity (\ref{Eq: SWAP})) one can show that (\ref{Eq: ACA}) vanishes identically.  
This allows to rewrite (\ref{Eq: W_expansion}) as
\begin{equation} \label{Eq: mathcal_p3}
\fl \vec{N}_-^{(t)}=\Phi^{(t)}\vec{N}_-^{(0)}\\
=\left(A^t + p \sum_{n=1}^t A^{t-n} B A^n+p^2 \sum_{n=1}^{t-1}\sum_{m>n}^t A^{t-m}BA^n B A^{m-n}\right) \vec{N}_-^{(0)} + \mathcal{O}(p^3).
\end{equation} 
At variance with the corrections (\ref{Eq: C_term_1}),(\ref{Eq: C_term_2}),(\ref{Eq: C_term_3}), the $\mathcal{O}(p^2)$ corrections in (\ref{Eq: mathcal_p3}) correspond to freezing two regions at \textit{two} distinct times.  To see this, let us denote $(*)=\sum_{n=1}^{t-1}\sum_{m>n}^t\langle (A^{t-m}BA^{n}BA^{m-n})\vec{N}_-^{(0)}\rangle_{\rm LR}$ , where the LR subscript indicates that the e.v. is evaluated within linear response -- see below. Proceeding identically as before we get
\begin{eqnarray}
\fl (*)=\sum_{n=1}^{t-1}\sum_{m>n}^t\sum_{i,j=1}^L\langle (A^{t-m}(B_{0,i}+B_{1,i}) A^{n}(B_{0,j}+B_{1,j})A^{m-n}) \vec{N}_{-}^{(0)}\rangle_{\rm LR} \nonumber\\
\fl \cong \sum_{n=1}^{t-1}(t-n)\sum_{i,j,l=1}^L   \left( \langle \hat{\Pi}_i (\hat{F}^\dagger)^n\hat{\Pi}_j \hat{\rho}_{-,l}\hat{\Pi}_j\hat{F}^n\hat{\Pi}_i\rangle_{\rm LR} \right. \nonumber -\langle \hat{\Pi}_i (\hat{F}^\dagger)^n\hat{\Theta}_j \hat{\rho}_{-,l}\hat{\Theta}_j\hat{F}^n\hat{\Pi}_i\rangle_{\rm LR}  \nonumber \\
-\langle \hat{\Theta}_i (\hat{F}^\dagger)^n\hat{\Pi}_j \hat{\rho}_{-,l}\hat{\Pi}_j \hat{F}^n\hat{\Theta}_i \rangle_{\rm LR} \nonumber \left.+\langle \hat{\Theta}_i (\hat{F}^\dagger)^n\hat{\Theta}_j \hat{\rho}_{-,l}\hat{\Theta}_j\hat{F}^n\hat{\Theta}_i\rangle_{\rm LR}\right).
\end{eqnarray}
As always we have made use of conservation of the global charges under the nonperturbed dynamics $[\hat{F},\hat{N}_{+,-}]=0$.  This equation has a simple physical interpretation as alluded earlier. It gives us the contribution to the decay of the global charge $N_-$ stemming from two \textit{frozen} regions separated in time by $n$ steps.  We can simplify this expression further 
\begin{equation} \label{Eq: star_equals}
\fl (*)\cong 2\sum_{n=1}^t (t-n)\sum_{i,j=1}^L \left(\langle \hat{\Pi}_i (\hat{F}^\dagger)^n\hat{\Pi}_j\hat{\rho}_{-,j}\hat{\Pi}_j\hat{F}^n\hat{\Pi}_i\rangle_{\rm LR} -\langle \hat{\Theta}_i (\hat{F}^\dagger)^n \hat{\Pi}_j \hat{\rho}_{-,j} \hat{\Pi}_j \hat{F}^n \hat{\Theta}_i \rangle_{\rm LR} \right),
\end{equation}
where once again we have used (\ref{Eq: SWAP}). 
\begin{figure} 
\centering 
\includegraphics[scale=0.5]{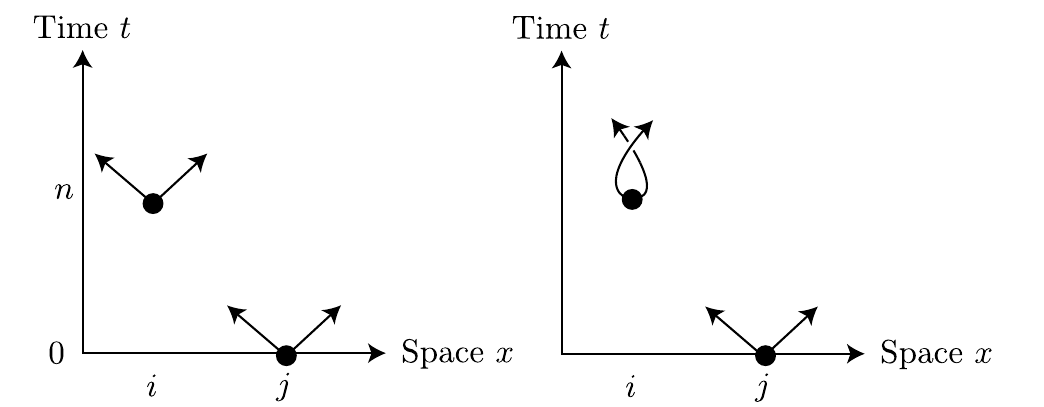}
\caption{Processes contributing to (\ref{Eq: star_equals}).}
\label{Fig: pic3}
\end{figure}
The two types of terms contributing to (\ref{Eq: star_equals}) are depicted in Fig. \ref{Fig: pic3}. We can now give an even clearer interpretation of this equation within linear response theory.  Let the linear response (LR) background state be given by $\hat{\rho}_{\rm LR} \propto \exp(\mu \hat{N}_+ - \delta \hat{N}_-)$, with $|\delta| \ll 1$ giving rise to a small imbalance in the number of right movers vs left movers.  Expanding the state $\hat{\rho}_{\rm LR} \propto \exp(\mu \hat{N}_+)(1-\delta \hat{N}_-)+\mathcal{O}(\delta ^2)$ and using (\ref{Eq: SWAP}) yields 
\begin{equation}
\fl \langle \hat{\Pi}_i (\hat{F}^\dagger)^n\hat{\Pi}_{-,j}\hat{F}^n\hat{\Pi}_i\rangle_{\rm LR} -\langle \hat{\Theta}_i (\hat{F}^\dagger)^n \hat{\Pi}_{-,j} \hat{F}^n \hat{\Theta}_i \rangle_{\rm LR} = \\
=-2\delta \langle (\hat{F}^\dagger)^n \hat{\Pi}_{-,j} \hat{F}^n \hat{\Pi}_{-,i} \rangle +\mathcal{O}(\delta ^2),
\end{equation}
where the expectation values are now w.r.t. the unperturbed background state $\langle \cdots \rangle \equiv \frac{1}{\tr e^{\mu \hat{N}_+}}\tr(\cdots e^{\mu \hat{N}_+})$.  We have also made use of the trivial relation $[\hat{\Theta}_i,\hat{N}_+]=0$, which is just a restatement that $\hat{N_+}$ is conserved under the full dynamics. Let us refer to $\hat{\Pi}_{-,j}$ as the \textit{diffusive movers} which are nothing but eigenoperators of the Kraus map with eigenvalue $(1-2p)$, i.e. $\hat{K}_{\vec{\mu}}\hat{\Pi}_{-,i}\hat{K}_{\vec{\mu}}=(1-2p)\hat{\Pi}_{-,i}$. In other words, its orthogonal complement, $(1-\hat{\Pi}_i)\hat{\rho}_{-,i}$, is not affected by dissipation: $\hat{K}_{\vec{\mu}}(1-\hat{\Pi}_i)\hat{\rho}_{-,i}\hat{K}_{\vec{\mu}}=(1-\hat{\Pi}_i)\hat{\rho}_{-,i}$. Using translation invariance we arrive at our final expression
\begin{equation} \label{Eq: final_2nd_order}
(*)\cong -4 \delta \sum_{n=1}^{t-1} (t-n) \langle \hat{\Pi}^{0(n)}_- \hat{\Pi}_{-}\rangle^c+ \mathcal{O}(\delta ^2),
\end{equation}
where we have used the freedom to take instead the connected $2$-point function since $\langle \hat{\Pi}_-\rangle=0$ from symmetry arguments. This equation has a very natural meaning. The second order corrections are given in terms of the correlator $\frac{1}{L}\langle \hat{\Pi}_-^{0(n)} \hat{\Pi}_-\rangle^c \equiv \sum_i \langle \hat{\Pi}_{-,L/2}^{0(n)} \hat{\Pi}_{-,i}\rangle^c$, which gives us the overlap between the diffusive movers $\hat{\Pi}_{-,i}$ at different times when evolved under the unperturbed dynamics.  From here it follows that 
\begin{equation}
\fl \langle \hat{N}_-^{(t)}\hat{N}_-^{(0)} \rangle^c = \left(1-2pt\gamma_{s.t.}+4p^2\frac{1}{LG_{+,+}}\sum_{n=1}^{t-1}(t-n)\langle \hat{\Pi}_-^{0(n)}\hat{\Pi}_-\rangle^c \right)\langle \hat{N}_-^2 \rangle^c +\mathcal{O}(p^3)
\end{equation}
The fact that $\langle \hat{\Pi}_-^{0(t)}\hat{\Pi}_-\rangle^c$ saturates at short enough times to a positive constant (see Fig. \ref{Fig: Pi_corr}) indicates that the two point function of the imbalance of solitons should scale as $\langle \hat{N}_-^{(t)}\hat{N}_-^{(0)}\rangle^c \to (1-2pt\gamma_{\rm s.t.}+4(pt)^2\tilde{\gamma}^2)\langle \hat{N}_-^2\rangle^c + \mathcal{O}(p^3)$ as $t \to \infty$, where we have defined $\tilde{\gamma}^2\equiv \tilde{D}(\mu)/2G_{+,+}$ with
\begin{equation}
\tilde{D}(\mu)\equiv \lim_{t\to\infty} \frac{1}{L} \langle \hat{\Pi}_-^{0(t)} \hat{\Pi}_-\rangle^c.
\end{equation} 
We find that $\tilde{\gamma}(\mu \to \infty) = 1$, just as $\gamma_{\rm s.t.}$. This is consistent with our expectations, since at low filling particles should decay exponentially fast with a decay rate given by $\Gamma=2p$ (see beginning of 
Sec. \ref{sec_tracer}). Further, we find numerically that $\tilde{\gamma}>\gamma_{\rm s.t.}$ which indicates that the effective decay rate at $t=1$ is actually smaller than that at $t=0$, which is consistent with our numerical results - see Fig. \ref{fig_JpJp_MPO}.
The analysis at this order in perturbation theory already reveals much information.  In the scaling limit, $p \to 0^+$, $t\to \infty$, $tp<\infty$, one finds that the entire perturbative series of $\Phi^{(t)}$ is solely determined by the operators $A$ (giving the $\hat{F}$ gates) and $B$ (giving the first order corrections in $p$ to the Kraus map $\sum_{\{\vec{\mu}\}} \hat{K}_{\vec{\mu}}\otimes \hat{K}_{\vec{\mu}}$). That is 
\begin{equation} \label{Eq: Phi_map}
\Phi^{(t)}=\left( \sum_{\{\vec{\mu}\}} \hat{K}_{\vec{\mu}}\hat{F}^\dagger \otimes \hat{K}_{\vec{\mu}} \hat{F}^\dagger\right)^t \asymp ((\mathbb{1}+pB)A)^t,
\end{equation}
 with $\asymp$ denoting the scaling limit. This is because at any order $n$ in $p$, the only terms contributing $t^n$ will be those containing precisely $n$ $B$'s, with the rest contributing at most $t^{n-1}$ (in the case of $n=2$ it happens that there are no order $tp^2$ terms).  Ultimately this means that the e.o.m.  for the negative movers in the scaling limit should be of the form
\begin{equation} \label{Eq: rho_minus_eom}
\vec{\rho}_-^{(t+1)}-\vec{\rho}_-^{(t)}=-\Gamma \vec{\rho}_-^{(t)},
\end{equation}
with the decay matrix given by $\Gamma=\mathbb{1}-A-pBA$.  From here it follows 
\begin{eqnarray}
\fl \partial_t \langle \hat{N}_-^{(t)}\hat{N}_-^{(0)}\rangle = p\sum_i \langle \hat{\Theta}_i \hat{F}^\dagger \hat{N}_-^{(t)}\hat{F}\hat{\Theta}_i\hat{N}_-^{(0)} \rangle - \langle \hat{\Pi}_i \hat{F}^\dagger \hat{N}_-^{(t)}\hat{F}\hat{\Pi}_i\hat{N}_-^{(0)} \rangle \nonumber \\
=-2p\sum_i \langle \hat{F}^{\dagger}\hat{N}_-^{(t)}\hat{F}\hat{\Pi}_i\hat{\rho}_{-,i}\rangle \nonumber\\
=-2p\langle \hat{N}_-^{(t)}\hat{\Pi}_-^{0(-1)}\rangle,
\end{eqnarray}
where in the second line we have used (\ref{Eq: SWAP}) and $[\hat{\Theta}_i,\hat{N}_+]=0$ $\forall i$, and in the third line we have used $[\hat{F},\hat{N}_+]=0$ to evolve the operator $\hat{\Pi}_-$ backwards by one time step under the unitary $\hat{F}$. From here we extract the decay rate 
\begin{equation} \label{eq_decay_rate_eff}
\Gamma(t) = 2p\gamma(pt), \hspace{0.1in} \gamma(pt)\equiv\frac{\langle \hat{N}_-^{(t)}\hat{\Pi}_-^{0(-1)} \rangle}{\langle \hat{N}_-^{(t)}\hat{N}_-^{(0)}\rangle}.
\end{equation}
We emphasize that a perturbative expansion in our setup is only controlled in the limit $tp \to 0^+$, which simply prohibits extracting an analytical decay rate at long enough times. 
In the limit $tp \gg 1$ the function $\Gamma(t)$ becomes time independent, so that the imbalance decays exponentially $\langle N_-(t) \rangle_{LR} \sim \exp(-2p\gamma t)$ at sufficiently long times, with a decay rate $\gamma$ being a function of the density of particles that seems to be well approximated by $\gamma=(1-n)\gamma^*$ (see Fig. \ref{Fig: tracer_vs_full} in the main text).  Such scenario would be consistent with thermalization,  despite the fact that the thermalization rate is not given by a simple FGR estimate, and does not control the diffusion constant -- which depends on the full function $\gamma(pt)$.

There is an alternative view to the effective decay rate (\ref{eq_decay_rate_eff}).  Applying (\ref{Eq: Phi_map}) to the total imbalance $\hat{N}_-$ and taking the e.v. within LR we find that at order $n$ in the perturbation strength $p$ we must evaluate terms of the form 
\begin{equation} \label{eq_matrix_element}
\fl (-1)^{\pi_{\vec{X}}}\langle (\hat{F}^\dagger)^{\alpha_1}\hat{X}_{i_1}(\hat{F}^{\dagger})^{\alpha_2}\hat{X}_{i_2}\cdots (\hat{F}^\dagger)^{\alpha_n}\hat{X}_{i_n}(\hat{F}^\dagger)^{\alpha_{n+1}}\hat{N}_-\hat{F}^{\alpha_{n+1}}\hat{X}_{i_n}\hat{F}^{\alpha_n}\cdots \hat{X}_{i_2}\hat{F}^{\alpha_2}\hat{X}_{i_1}\hat{F}^{\alpha_1}\rangle_{\rm LR},
\end{equation}
where $\alpha_i$ are nonnegative integers with the constraint $\alpha_1+\alpha_2+\cdots+\alpha_{n+1}=t$, $i_j \in \{1,\cdots,L\}$, $\hat{X} \in \{\hat{\Pi},\hat{\Theta}\}$, and $\pi_{\vec{X}}=1$ if there are an odd number of $\hat{\Pi}$ operators in the array $\vec{X}=(\hat{X}_{i_1},\hat{X}_{i_2},\cdots,\hat{X}_{i_n})$, and $\pi_{\vec{X}}=0$ otherwise.  More precisely, the contribution to $\langle \hat{N}^{(t)}\rangle_{\rm LR}$ at order $n$ corresponds to the sum of (\ref{eq_matrix_element}) over all possible time indices $\{\alpha_i\}$ fulfilling the above constraint, spatial indices $\{i_j\}$, and operators $\{\hat{X}_{i_j}\}$. The upshot of this is that the effective decay rate at time $t$ will be governed by all possible processes of the form (\ref{eq_matrix_element}), where $n=1,2,\cdots, t$. As such, the effective decay rate is no longer controlled by single processes acting within some time window $\Delta t=1$ (one single time step), but rather by the set of all possible processes spanning time intervals $\Delta t=1,2,\cdots, t$.  In other words, processes involving $\Delta t >1$ cannot in general be factored into products of single processes (as has been shown explicitly for $n=2$).

\begin{figure}
\centering
\includegraphics[scale=0.26]{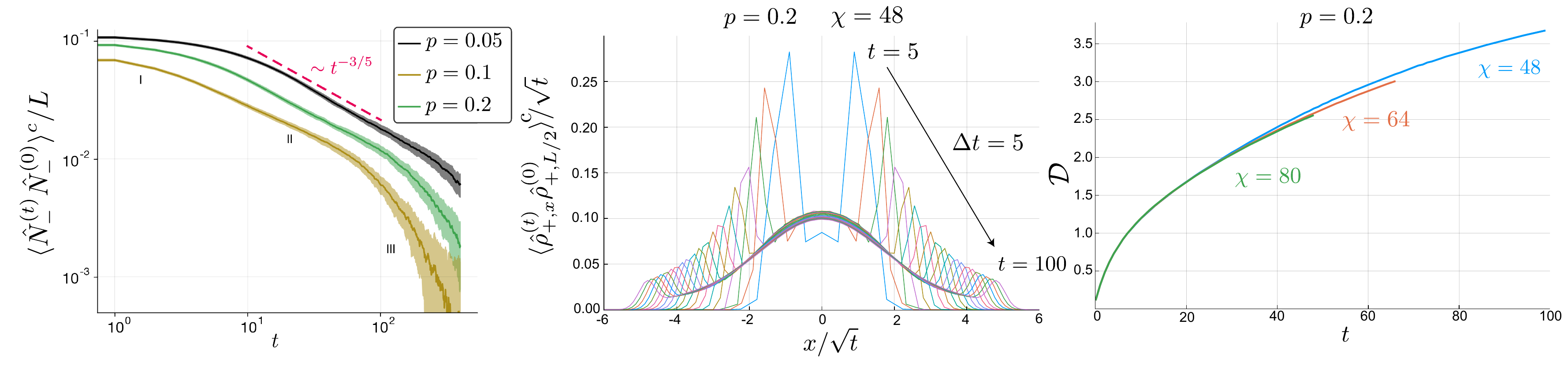}
\caption{\textbf{Transport in the noisy Rule 54 CA.} Results shown for half filling $\mu=0$. Left: Decay of imbalance for various values of strength of perturbation $p$.  After a brief period of exponential decay (I) the system goes through a transient regime of anomalous decay (II) followed by an exponential relaxation again in the longest time scales (III). Only regimes (I) and (III) are captured by the theory.  Results from MC. Middle: structure factor of density of movers reflecting the anomalously slow decay of imbalance; the satellite peaks fail to dampen.  Results using tMPO with $\chi=48$. Right: Time dependent diffusion constant extracted from the time derivative of the variance of the structure factor, showing some apparent ``superdiffusive'' regime. Results using tMPO for various $\chi$.}
\label{fig_transport_mu_0}
\end{figure}

In Fig. \ref{fig_transport_mu_0} we show the time evolution of the correlator $\langle \hat{N}_-^{(t)}\hat{N}_-^{(0)}\rangle$.  We can distinguish three regimes.  The dynamics at short times is well approximated by (\ref{Eq: N_-_derivative_1st_order}) when neglecting higher order terms in $p$.  (Recall that strictly speaking, Eq. (\ref{Eq: N_-_derivative_1st_order}) is only valid at $t=0$ where the decay rate is given by $\gamma_{\rm s.t. }$. ) This short-time, fast decay regime is followed by a long transient of anomalous decay. At the longest time scales,  the charge relaxes exponentially, but with a decay rate that is different from that obtained from (\ref{Eq: N_-_derivative_1st_order}) when neglecting $\mathcal{O}(p^2)$ terms.  The physical mechanism behind this rather slow relaxation can be captured when computing the structure factor $\langle \hat{\rho}_{+,x}^{(t)}\hat{\rho}_{+,x}^{(0)} \rangle^{\rm c}$ shown in the middle panel in Fig. \ref{fig_transport_mu_0}. While the central region develops a bump collapsing to a gaussian signaling diffusion, the ballistic peaks fail to disappear. This interplay between diffusive and ballistic spreading lead to some apparent superdiffusive behavior of the current-current correlator (alternatively, the charge $\hat{N}_-$), see right panel of Fig. \ref{fig_transport_mu_0}.

\subsection{3rd order corrections}
\begin{figure}
\centering
\includegraphics[scale=0.31]{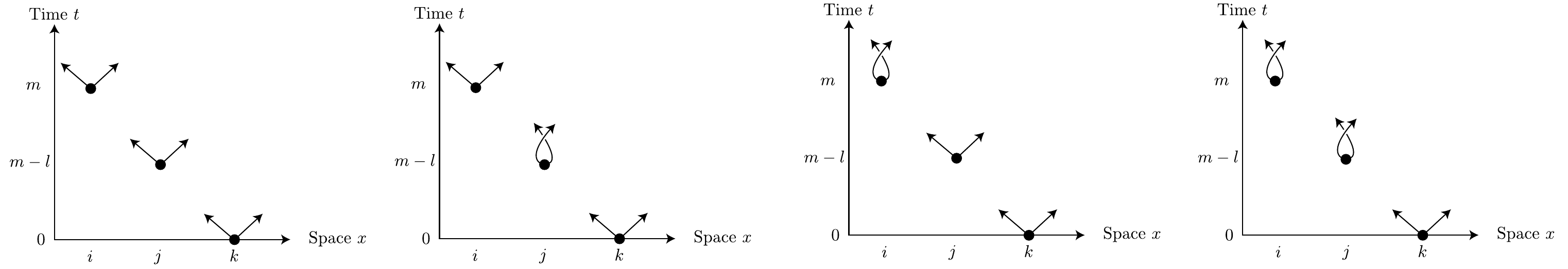}
\caption{Processes contributing to the third order corrections, see Eq. (\ref{Eq: 3rd_order_corrs})}
\end{figure}
\begin{figure}
\centering
\includegraphics[scale=0.23]{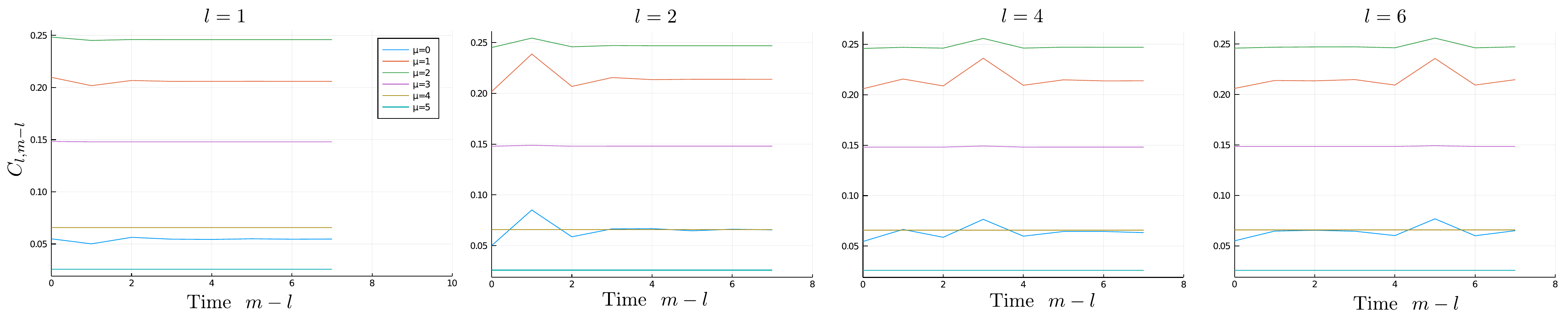}
\caption{Correlators $C_{l,m-l}$.}
\label{Fig: C_lm_corrs}
\end{figure}

Going one order higher up in the perturbative expansion we find 
\begin{eqnarray}
\fl \vec{N}_-^{(t)}=\Phi^{(t)}\vec{N}_-^{(0)} \nonumber \\
\fl\asymp\left(A^t + p \sum_{n=1}^t A^{t-n} B A^n+p^2 \sum_{n=1}^{t-1}\sum_{m>n}^t A^{t-m}BA^n B A^{m-n} + \right.  \\ 
\fl \hspace{1in} \left. + p^3 \sum_{l=1}^{t-2}\sum_{m>l}^{t-1}\sum_{n>m}^t A^{t-n}BA^{l}BA^{m-l}BA^{n-m} \right) \vec{N}_-^{(0)} + \mathcal{O}(p^4). \nonumber
\end{eqnarray} 
Recalling $B=\sum_{i=1}^L B_{0,i}+B_{1,i}$, and taking the expectation value of the last term in the previous expression in an homogeneous state within LR we get $8$ terms. Let us refer $(**) \equiv  \langle \left(\sum_{l=1}^{t-2}\sum_{m>l}^{t-1}\sum_{n>m}^t A^{t-n}BA^{l}BA^{m-l}BA^{n-m} \right) \vec{N}_-^{(0)}\rangle_{\rm LR}$. Using (\ref{Eq: SWAP}) together with $[\hat{F},\hat{N}_{+,-}]=0$ we find that only $4$ terms contribute 
\begin{eqnarray} \label{Eq: 3rd_order_corrs}
\fl (**)\cong -2 \sum_{l=1}^{t-2} \sum_{m>l}^{t-1} (t-m) \sum_{i,j,k=1}^L \left( \langle \hat{\Pi}_i (\hat{F}^\dagger)^l \hat{\Pi}_j \hat{\Pi}_{-,k}^{0(m-l)}  \hat{\Pi}_j \hat{F}^l \hat{\Pi}_i \rangle_{\rm LR} - \langle \hat{\Pi}_i (\hat{F}^\dagger)^l \hat{\Theta}_j  \hat{\Pi}_{-,k}^{0(m-l)}   \hat{\Theta}_j \hat{F}^l \hat{\Pi}_i \rangle_{\rm LR} \right. \nonumber\\
\fl \left.- \langle \hat{\Theta}_i (\hat{F}^\dagger)^l \hat{\Pi}_j \hat{\Pi}_{-,k}^{0(m-l)}  \hat{\Pi}_j \hat{F}^l \hat{\Theta}_i \rangle_{\rm LR} +\langle \hat{\Theta}_i (\hat{F}^\dagger)^l \hat{\Theta}_j \hat{\Pi}_{-,k}^{0(m-l)} \hat{\Theta}_j \hat{F}^l \hat{\Theta}_i \rangle_{\rm LR} \right) .
\end{eqnarray}
Simplifying using (\ref{Eq: SWAP})
\begin{eqnarray}
\fl (**) \cong 4 \delta \sum_{l=1}^{t-2}\sum_{m>l}^{t-1}(t-m)\sum_{i,j,k=1}^L \left( \langle (\hat{F}^\dagger)^l \hat{\Pi}_j \hat{\Pi}_{-,k}^{0 (m-l)}\hat{\Pi}_j \hat{F}^l \hat{\Pi}_{-,i} \rangle - \langle (\hat{F}^\dagger)^l \hat{\Theta}_j\hat{\Pi}_{-,k}^{0(m-l)} \hat{\Theta}_j \hat{F}^l \hat{\Pi}_{-,i} \rangle \right).\nonumber
\end{eqnarray}
We can improve slightly this expression invoking translation invariance as usual
\begin{equation}
\fl (**) \cong 4 L \delta \sum_{l=1}^{t-2}\sum_{m>l}^{t-1}(t-m)\sum_{i,j=1}^L \left( \langle (\hat{F}^\dagger)^l \hat{\Pi}_j \hat{\Pi}_{-,L/2}^{0 (m-l)}\hat{\Pi}_j \hat{F}^l \hat{\Pi}_{-,i} \rangle - \langle (\hat{F}^\dagger)^l \hat{\Theta}_j \hat{\Pi}_{-,L/2}^{0(m-l)} \hat{\Theta}_j \hat{F}^l \hat{\Pi}_{-,i} \rangle \right).
\end{equation}
The task now is thus to evaluate the $3-point$ function 
\begin{equation}
C_{l,m-l} \equiv \sum_{i,j=1}^L\langle (\hat{F}^\dagger)^l \hat{\Pi}_j \hat{\Pi}_{-,L/2}^{0 (m-l)}\hat{\Pi}_j \hat{F}^l \hat{\Pi}_{-,i} \rangle - \langle (\hat{F}^\dagger)^l \hat{\Theta}_j\hat{\Pi}_{-,L/2}^{0(m-l)} \hat{\Theta}_j \hat{F}^l \hat{\Pi}_{-,i} \rangle.
\end{equation}
First note we can rewrite this as 
\begin{equation} \label{Eq: corr_C}
C_{l,m-l} = \sum_{j=1}^L\langle \hat{\Pi}_{-,i}^{0(-l)} \hat{\Pi}_j \hat{\Pi}_{-,L/2}^{0 (m-l)}\hat{\Pi}_j \rangle - \langle \hat{\Pi}_{-,i}^{0(-l)} \hat{\Theta}_j\hat{\Pi}_{-,L/2}^{0(m-l)} \hat{\Theta}_j \rangle.
\end{equation}
The results of evaluating $C_{l,m-l}$ via tMPO are shown in Fig. (\ref{Fig: C_lm_corrs}) showing also plateaus, indicating that the third order corrections should be taken into account as well.  
\section{Numerical details}
\subsection{MPO calculations} \label{Sec: MPO_calculations}
\begin{figure}[h!]
    \centering
    \includegraphics[scale=0.33]{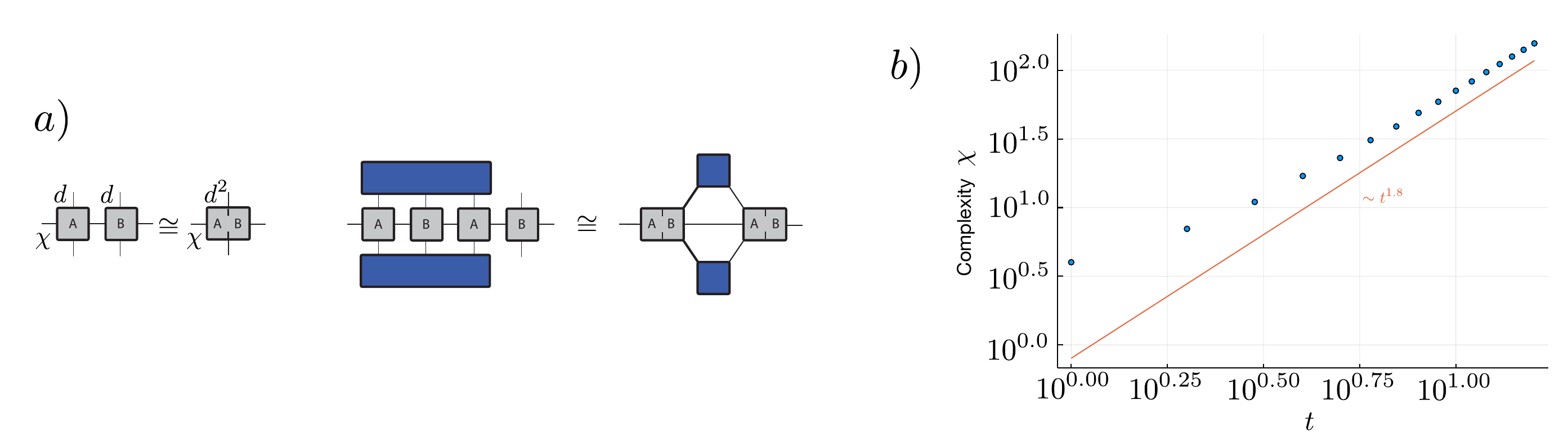}
    \caption{\textbf{MPO notation and complexity growth in Rule 54.} a)Left:  Given some initial state $\hat{\rho}$ in MPO form with \textit{blocks} (tensors) of size $\chi \times d \times d \times \chi$ with bond dimension $\chi$ and local Hilbert space dimension $d$ ($d=2$ in our case) we merge each pair of sites within a unit cell to form a new tensor of size $\chi \times d^2 \times d^2 \times \chi$. Right: The unitary gates in the Rule 54 model act on three sites. By decomposing the system into unit cells each gate effectively acts on two blocks. Double (single) sized lines denote the gate is acting on the two (leftmost of the) sites in each unit cell. b) Complexity growth in the Rule 54 vs time for an initial operator (here shown for $\hat{\mathcal{O}}^{(0)}=\hat{J}_{-,L/2}$). Results for the maximum bond dimension of $\hat{\mathcal{O}}^{(t)}$ as a function of time which grows roughly as $t^2$, consistent with \cite{klobas2019time}, as opposed to exponentially in time.}
    \label{Fig: notation_q_channel}
\end{figure}

The implementation of the MPO time evolution of the quantum channel described in the main text is straightforward. Since both the unperturbed (Rule 54) and full dynamics (including the Kraus operators) contains at most three site gates (excluding projectors) we construct our Hilbert space in terms of unit cells. This means all gates we will be dealing with will act effectively on two sites (see Fig. \ref{Fig: notation_q_channel}). Each time step comprises two layers given in terms of the Rule 54 gates followed by one given in terms of the set Kraus operators described in the main text. To simulate this we \textit{vectorize} the channel so that the initial state is stored as a matrix product state (MPS) of local Hilbert space dimension $d^4$ ($d=2$ for our qubit system) and proceed forward as in standard time-evolving block decimation (TEBD) based routines. Note that the layer of Kraus operators involves sites that are acted upon simultaneously by three Kraus operators but that is fine since these mutually commute. A small technical artifact of using this approach is that applying the layer of Kraus operators necessarily entails adding two or more MPSs resulting on a new MPS with bond dimension equal to the sum of each of the bond dimensions of the MPSs. This is because $\hat{K}_{\mu_i=1}\otimes \hat{K}_{\mu_i=1}+\hat{K}_{\mu_i=2}\otimes \hat{K}_{\mu_i=2}$ cannot be put onto a two-site gate form (since it involves three sites [five sites in the original Hilbert space]). To get around this issue we first add the MPSs in a local fashion e.g.  
\begin{eqnarray*}
\fl |\hat{\rho})_{\rm NEW}=A_{i-1,i,i+1}|\hat{\rho})+B_{i-1,i,i+1}|\hat{\rho})=\\
\fl=\sum_{\underline{\sigma}}\tr[\textbf{M}^{\sigma_1}...\textbf{M}^{\sigma_{i-2}}(\matrix{\widetilde{\textbf{M}}^{\sigma_{i-1}} & \widetilde{\widetilde{\textbf{M}}}^{\sigma_{i-1}}})\left(\matrix{ \widetilde{\textbf{M}}^{\sigma_{i}} & 0 \cr 0 & \widetilde{\widetilde{\textbf{M}}}^{\sigma_{i}} }\right)\left(\matrix{\widetilde{\textbf{M}}^{\sigma_{i+1}} \cr \widetilde{\widetilde{\textbf{M}}}^{\sigma_{i+1}}}\right)\textbf{M}^{\sigma_{i+2}}...\textbf{M}^{\sigma_N}]|\sigma_1...\sigma_N)
\end{eqnarray*}
resulting in a bond dimension growth on only those three sites. Since the Kraus operators involve mostly projectors we can fix the maximum bond dimension of the resulting MPS to be the maximum of each of the two involved plus an extra small bond dimension, i.e. $\chi_{\rm NEW}=\rm max(\tilde{\chi},\tilde{\tilde{\chi}})+\delta$, with $\delta = \mathcal{O}(1)$. Thus we have two sources of truncation error during the entire quantum channel: 1) errors due to truncation from applying the unitary gates and, 2) errors from truncation after adding $\geq$2 MPSs.  All tensor network contractions are implemented using the Julia package ITensors.jl \cite{fishman2020itensor}.

\subsection{Monte Carlo calculations}
The MPO based approach described above works very well when we limit ourselves to short to intermediate time dynamics, which for many instances it is fine. In the particular case of the dissipative Rule 54 model and away from the low density limit we find that the dynamics of the non-conserved charge $N_-$ decays very slowly before it relaxes at long enough times. To have access to this large time regime we make use of the Metropolis algorithm to compute the observables of interest for a given ensemble (given by fixed $\mu \equiv \mu_+$). We are mostly interested in computing the (connected) two-point functions $\langle \hat{N}_-^{(t)}\hat{N}_-^{(0)} \rangle^c$. We remark once again that the behavior of the current-current correlator, $\langle \hat{J}_+^{(t)}\hat{J}_+^{(0)} \rangle^c$, is very similar to that of $\langle \hat{N}_-^{(t)}\hat{N}_-^{(0)} \rangle^c$ (this is because $\hat{J}_+^{(t)}=\hat{N}_-^{(t+1)}$ for $p>0$). For the times involved in this work $t \sim \mathcal{O}(10^2)$ and the system sizes considered $L \sim \mathcal{O}(10^2)-\mathcal{O}(10^3)$ we have required of the order of $\mathcal{O}(10^7)$ realizations which we compartmentalize into approximately $20$ \textit{bins} that we later use for error analysis. We remark that the MC equilibration time takes longer as we move away from the half-filled limit.  

The above analysis concerns transport in the noisy Rule 54 CA. The MC analysis for tracer dynamics is somewhat different. In particular,  tracer dynamics is more susceptible to the strength of the perturbation than when studying transport. While for transport we already find collapsed results using $p \sim \mathcal{O}(0.1)$,  for tracer dynamics we must consider at most $p\sim \mathcal{O}(0.01)$ (so that our numerical results are in accordance with our theory predictions).  This means that the crossover to diffusion in the tracer distribution happens at much longer times. In this regard it is worth emphasizing that sampling the tracer distribution $\omega(x,t)$ (giving us access to the self-diffusion constant $\mathcal{D}^*$) is much simpler than the full many-body correlator $\langle \hat{\rho}_{+,x}^{(t)}\hat{\rho}^{(0)}_{+,L/2}\rangle$ (giving us access to the diffusion constant $\mathcal{D}$). By this we mean: $1)$ The number of sampling iterations is much greater in the latter case. Only $\mathcal{O}(10^3)$ samples are needed to sample faithfully the tracer distribution for the accessible times $t \sim \mathcal{O}(10^3)-\mathcal{O}(10^4)$. $2)$ The tracer distribution is not contaminated as much by finite size effects. It is clear that the maximum accessible time scale to probe the many-body correlator is determined by the Lieb-Robinson velocity. Given that in Rule 54 this is $\mathcal{O}(1)$ we are bounded to time scales of the order of the system size. At variance, when probing the tracer distribution we can exploit the fact that we are measuring the number of solitons that traveled a given \textit{distance} away from their starting points.  We can exploit this to our advantage and set much smaller system sizes $L \ll t_{\rm max}$ while imposing pbc. For our tracer distribution simulations we never exceed $L=1200$, while the maximum times considered go past $t\sim 10000$. 

\bibliographystyle{IEEEtran}
\bibliography{refs}

\begin{thebibliography}{10}
\providecommand{\url}[1]{#1}
\csname url@samestyle\endcsname
\providecommand{\newblock}{\relax}
\providecommand{\bibinfo}[2]{#2}
\providecommand{\BIBentrySTDinterwordspacing}{\spaceskip=0pt\relax}
\providecommand{\BIBentryALTinterwordstretchfactor}{4}
\providecommand{\BIBentryALTinterwordspacing}{\spaceskip=\fontdimen2\font plus
\BIBentryALTinterwordstretchfactor\fontdimen3\font minus
  \fontdimen4\font\relax}
\providecommand{\BIBforeignlanguage}[2]{{%
\expandafter\ifx\csname l@#1\endcsname\relax
\typeout{** WARNING: IEEEtran.bst: No hyphenation pattern has been}%
\typeout{** loaded for the language `#1'. Using the pattern for}%
\typeout{** the default language instead.}%
\else
\language=\csname l@#1\endcsname
\fi
#2}}
\providecommand{\BIBdecl}{\relax}
\BIBdecl

\bibitem{spohn2012large}
H.~Spohn, \emph{Large scale dynamics of interacting particles}.\hskip 1em plus
  0.5em minus 0.4em\relax Springer Science \& Business Media, 2012.

\bibitem{rothman2004lattice}
D.~H. Rothman and S.~Zaleski, \emph{Lattice-gas cellular automata: simple
  models of complex hydrodynamics}.\hskip 1em plus 0.5em minus 0.4em\relax
  Cambridge University Press, 2004, vol.~5.

\bibitem{hardy1973time}
J.~Hardy, Y.~Pomeau, and O.~De~Pazzis, ``Time evolution of a two-dimensional
  model system. i. invariant states and time correlation functions,''
  \emph{Journal of Mathematical Physics}, vol.~14, no.~12, pp. 1746--1759,
  1973.

\bibitem{PhysRevA.13.1949}
\BIBentryALTinterwordspacing
J.~Hardy, O.~de~Pazzis, and Y.~Pomeau, ``Molecular dynamics of a classical
  lattice gas: Transport properties and time correlation functions,''
  \emph{Phys. Rev. A}, vol.~13, pp. 1949--1961, May 1976. [Online]. Available:
  \url{https://link.aps.org/doi/10.1103/PhysRevA.13.1949}
\BIBentrySTDinterwordspacing

\bibitem{PhysRevLett.56.1505}
\BIBentryALTinterwordspacing
U.~Frisch, B.~Hasslacher, and Y.~Pomeau, ``Lattice-gas automata for the
  navier-stokes equation,'' \emph{Phys. Rev. Lett.}, vol.~56, pp. 1505--1508,
  Apr 1986. [Online]. Available:
  \url{https://link.aps.org/doi/10.1103/PhysRevLett.56.1505}
\BIBentrySTDinterwordspacing

\bibitem{wolfram1986cellular}
S.~Wolfram, ``Cellular automaton fluids 1: Basic theory,'' \emph{Journal of
  statistical physics}, vol.~45, no.~3, pp. 471--526, 1986.

\bibitem{demasi1989hydrodynamics}
A.~DeMasi, R.~Esposito, J.~Lebowitz, and E.~Presutti, ``Hydrodynamics of
  stochastic cellular automata,'' \emph{Communications in mathematical
  physics}, vol. 125, no.~1, pp. 127--145, 1989.

\bibitem{PhysRevLett.119.110603}
\BIBentryALTinterwordspacing
M.~Medenjak, K.~Klobas, and T.~c.~v. Prosen, ``Diffusion in deterministic
  interacting lattice systems,'' \emph{Phys. Rev. Lett.}, vol. 119, p. 110603,
  Sep 2017. [Online]. Available:
  \url{https://link.aps.org/doi/10.1103/PhysRevLett.119.110603}
\BIBentrySTDinterwordspacing

\bibitem{gopalakrishnan2018facilitated}
S.~Gopalakrishnan and B.~Zakirov, ``Facilitated quantum cellular automata as
  simple models with non-thermal eigenstates and dynamics,'' \emph{Quantum
  Science and Technology}, vol.~3, no.~4, p. 044004, 2018.

\bibitem{iaconis2019anomalous}
J.~Iaconis, S.~Vijay, and R.~Nandkishore, ``Anomalous subdiffusion from
  subsystem symmetries,'' \emph{Physical Review B}, vol. 100, no.~21, p.
  214301, 2019.

\bibitem{iadecola2020nonergodic}
T.~Iadecola and S.~Vijay, ``Nonergodic quantum dynamics from deformations of
  classical cellular automata,'' \emph{Physical Review B}, vol. 102, no.~18, p.
  180302, 2020.

\bibitem{feldmeier2020anomalous}
J.~Feldmeier, P.~Sala, G.~De~Tomasi, F.~Pollmann, and M.~Knap, ``Anomalous
  diffusion in dipole-and higher-moment-conserving systems,'' \emph{Physical
  Review Letters}, vol. 125, no.~24, p. 245303, 2020.

\bibitem{iaconis2021multipole}
J.~Iaconis, A.~Lucas, and R.~Nandkishore, ``Multipole conservation laws and
  subdiffusion in any dimension,'' \emph{Physical Review E}, vol. 103, no.~2,
  p. 022142, 2021.

\bibitem{pozsgay2021yang}
B.~Pozsgay, ``A yang-baxter integrable cellular automaton with a four site
  update rule,'' \emph{arXiv preprint arXiv:2106.00696}, 2021.

\bibitem{gombor2021integrable}
T.~Gombor and B.~Pozsgay, ``Integrable spin chains and cellular automata with
  medium-range interaction,'' \emph{Physical Review E}, vol. 104, no.~5, p.
  054123, 2021.

\bibitem{gombor2021superintegrable}
------, ``Superintegrable cellular automata and dual unitary gates from
  yang-baxter maps,'' \emph{arXiv preprint arXiv:2112.01854}, 2021.

\bibitem{prosen2021reversible}
T.~Prosen, ``Reversible cellular automata as integrable interactions
  round-a-face: Deterministic, stochastic, and quantized,'' \emph{arXiv
  preprint arXiv:2106.01292}, 2021.

\bibitem{prosen2021many}
------, ``Many body quantum chaos and dual unitarity round-a-face,''
  \emph{arXiv preprint arXiv:2105.08022}, 2021.

\bibitem{prosen2016integrability}
T.~Prosen and C.~Mej{\'\i}a-Monasterio, ``Integrability of a deterministic
  cellular automaton driven by stochastic boundaries,'' \emph{Journal of
  Physics A: Mathematical and Theoretical}, vol.~49, no.~18, p. 185003, 2016.

\bibitem{klobas2019time}
K.~Klobas, M.~Medenjak, T.~Prosen, and M.~Vanicat, ``Time-dependent matrix
  product ansatz for interacting reversible dynamics,'' \emph{Communications in
  Mathematical Physics}, vol. 371, no.~2, pp. 651--688, 2019.

\bibitem{klobas2020matrix}
K.~Klobas, M.~Vanicat, J.~P. Garrahan, and T.~Prosen, ``Matrix product state of
  multi-time correlations,'' \emph{Journal of Physics A: Mathematical and
  Theoretical}, vol.~53, no.~33, p. 335001, 2020.

\bibitem{PhysRevE.102.062107}
\BIBentryALTinterwordspacing
J.~W.~P. Wilkinson, K.~Klobas, T.~c.~v. Prosen, and J.~P. Garrahan, ``Exact
  solution of the floquet-pxp cellular automaton,'' \emph{Phys. Rev. E}, vol.
  102, p. 062107, Dec 2020. [Online]. Available:
  \url{https://link.aps.org/doi/10.1103/PhysRevE.102.062107}
\BIBentrySTDinterwordspacing

\bibitem{klobas2021exact}
K.~Klobas and B.~Bertini, ``Exact relaxation to gibbs and non-equilibrium
  steady states in the quantum cellular automaton rule 54,'' \emph{SciPost
  Physics}, vol.~11, no.~6, p. 106, 2021.

\bibitem{gopalakrishnan2018operator}
S.~Gopalakrishnan, ``Operator growth and eigenstate entanglement in an
  interacting integrable floquet system,'' \emph{Physical Review B}, vol.~98,
  no.~6, p. 060302, 2018.

\bibitem{gopalakrishnan2018hydrodynamics}
S.~Gopalakrishnan, D.~A. Huse, V.~Khemani, and R.~Vasseur, ``Hydrodynamics of
  operator spreading and quasiparticle diffusion in interacting integrable
  systems,'' \emph{Physical Review B}, vol.~98, no.~22, p. 220303, 2018.

\bibitem{PhysRevLett.122.250603}
\BIBentryALTinterwordspacing
V.~Alba, J.~Dubail, and M.~Medenjak, ``Operator entanglement in interacting
  integrable quantum systems: The case of the rule 54 chain,'' \emph{Phys. Rev.
  Lett.}, vol. 122, p. 250603, Jun 2019. [Online]. Available:
  \url{https://link.aps.org/doi/10.1103/PhysRevLett.122.250603}
\BIBentrySTDinterwordspacing

\bibitem{klobas2021entanglement}
K.~Klobas and B.~Bertini, ``Entanglement dynamics in rule 54: exact results and
  quasiparticle picture,'' \emph{SciPost Physics}, vol.~11, no.~6, p. 107,
  2021.

\bibitem{PhysRevLett.123.170603}
\BIBentryALTinterwordspacing
A.~J. Friedman, S.~Gopalakrishnan, and R.~Vasseur, ``Integrable many-body
  quantum floquet-thouless pumps,'' \emph{Phys. Rev. Lett.}, vol. 123, p.
  170603, Oct 2019. [Online]. Available:
  \url{https://link.aps.org/doi/10.1103/PhysRevLett.123.170603}
\BIBentrySTDinterwordspacing

\bibitem{buvca2021rule}
B.~Bu{\v{c}}a, K.~Klobas, and T.~Prosen, ``Rule 54: Exactly solvable model of
  nonequilibrium statistical mechanics,'' \emph{arXiv preprint
  arXiv:2103.16543}, 2021.

\bibitem{PhysRevLett.126.160602}
\BIBentryALTinterwordspacing
K.~Klobas, B.~Bertini, and L.~Piroli, ``Exact thermalization dynamics in the
  ``rule 54'' quantum cellular automaton,'' \emph{Phys. Rev. Lett.}, vol. 126,
  p. 160602, Apr 2021. [Online]. Available:
  \url{https://link.aps.org/doi/10.1103/PhysRevLett.126.160602}
\BIBentrySTDinterwordspacing

\bibitem{PhysRevB.53.983}
\BIBentryALTinterwordspacing
X.~Zotos and P.~Prelov\ifmmode~\check{s}\else \v{s}\fi{}ek, ``Evidence for
  ideal insulating or conducting state in a one-dimensional integrable
  system,'' \emph{Phys. Rev. B}, vol.~53, pp. 983--986, Jan 1996. [Online].
  Available: \url{https://link.aps.org/doi/10.1103/PhysRevB.53.983}
\BIBentrySTDinterwordspacing

\bibitem{PhysRevB.76.245108}
\BIBentryALTinterwordspacing
P.~Jung and A.~Rosch, ``Spin conductivity in almost integrable spin chains,''
  \emph{Phys. Rev. B}, vol.~76, p. 245108, Dec 2007. [Online]. Available:
  \url{https://link.aps.org/doi/10.1103/PhysRevB.76.245108}
\BIBentrySTDinterwordspacing

\bibitem{PhysRevLett.110.070602}
\BIBentryALTinterwordspacing
M.~\ifmmode \check{Z}\else \v{Z}\fi{}nidari\ifmmode~\check{c}\else \v{c}\fi{},
  ``Coexistence of diffusive and ballistic transport in a simple spin ladder,''
  \emph{Phys. Rev. Lett.}, vol. 110, p. 070602, Feb 2013. [Online]. Available:
  \url{https://link.aps.org/doi/10.1103/PhysRevLett.110.070602}
\BIBentrySTDinterwordspacing

\bibitem{PhysRevB.91.115130}
\BIBentryALTinterwordspacing
C.~Karrasch, D.~M. Kennes, and F.~Heidrich-Meisner, ``Spin and thermal
  conductivity of quantum spin chains and ladders,'' \emph{Phys. Rev. B},
  vol.~91, p. 115130, Mar 2015. [Online]. Available:
  \url{https://link.aps.org/doi/10.1103/PhysRevB.91.115130}
\BIBentrySTDinterwordspacing

\bibitem{PhysRevB.90.094417}
\BIBentryALTinterwordspacing
R.~Steinigeweg, F.~Heidrich-Meisner, J.~Gemmer, K.~Michielsen, and H.~De~Raedt,
  ``Scaling of diffusion constants in the spin-$\frac{1}{2}$ xx ladder,''
  \emph{Phys. Rev. B}, vol.~90, p. 094417, Sep 2014. [Online]. Available:
  \url{https://link.aps.org/doi/10.1103/PhysRevB.90.094417}
\BIBentrySTDinterwordspacing

\bibitem{PhysRevB.93.205121}
\BIBentryALTinterwordspacing
A.~Biella, A.~De~Luca, J.~Viti, D.~Rossini, L.~Mazza, and R.~Fazio, ``Energy
  transport between two integrable spin chains,'' \emph{Phys. Rev. B}, vol.~93,
  p. 205121, May 2016. [Online]. Available:
  \url{https://link.aps.org/doi/10.1103/PhysRevB.93.205121}
\BIBentrySTDinterwordspacing

\bibitem{PhysRevLett.115.180601}
\BIBentryALTinterwordspacing
B.~Bertini, F.~H.~L. Essler, S.~Groha, and N.~J. Robinson, ``Prethermalization
  and thermalization in models with weak integrability breaking,'' \emph{Phys.
  Rev. Lett.}, vol. 115, p. 180601, Oct 2015. [Online]. Available:
  \url{https://link.aps.org/doi/10.1103/PhysRevLett.115.180601}
\BIBentrySTDinterwordspacing

\bibitem{PhysRevB.94.245117}
\BIBentryALTinterwordspacing
------, ``Thermalization and light cones in a model with weak integrability
  breaking,'' \emph{Phys. Rev. B}, vol.~94, p. 245117, Dec 2016. [Online].
  Available: \url{https://link.aps.org/doi/10.1103/PhysRevB.94.245117}
\BIBentrySTDinterwordspacing

\bibitem{PhysRevLett.85.1092}
\BIBentryALTinterwordspacing
A.~Rosch and N.~Andrei, ``Conductivity of a clean one-dimensional wire,''
  \emph{Phys. Rev. Lett.}, vol.~85, pp. 1092--1095, Jul 2000. [Online].
  Available: \url{https://link.aps.org/doi/10.1103/PhysRevLett.85.1092}
\BIBentrySTDinterwordspacing

\bibitem{PhysRevLett.103.216602}
\BIBentryALTinterwordspacing
J.~Sirker, R.~G. Pereira, and I.~Affleck, ``Diffusion and ballistic transport
  in one-dimensional quantum systems,'' \emph{Phys. Rev. Lett.}, vol. 103, p.
  216602, Nov 2009. [Online]. Available:
  \url{https://link.aps.org/doi/10.1103/PhysRevLett.103.216602}
\BIBentrySTDinterwordspacing

\bibitem{PhysRevB.83.035115}
\BIBentryALTinterwordspacing
------, ``Conservation laws, integrability, and transport in one-dimensional
  quantum systems,'' \emph{Phys. Rev. B}, vol.~83, p. 035115, Jan 2011.
  [Online]. Available:
  \url{https://link.aps.org/doi/10.1103/PhysRevB.83.035115}
\BIBentrySTDinterwordspacing

\bibitem{PhysRevB.88.115126}
\BIBentryALTinterwordspacing
Y.~Huang, C.~Karrasch, and J.~E. Moore, ``Scaling of electrical and thermal
  conductivities in an almost integrable chain,'' \emph{Phys. Rev. B}, vol.~88,
  p. 115126, Sep 2013. [Online]. Available:
  \url{https://link.aps.org/doi/10.1103/PhysRevB.88.115126}
\BIBentrySTDinterwordspacing

\bibitem{PhysRevLett.103.096402}
\BIBentryALTinterwordspacing
A.~Garg, D.~Rasch, E.~Shimshoni, and A.~Rosch, ``Large violation of the
  wiedemann-franz law in luttinger liquids,'' \emph{Phys. Rev. Lett.}, vol.
  103, p. 096402, Aug 2009. [Online]. Available:
  \url{https://link.aps.org/doi/10.1103/PhysRevLett.103.096402}
\BIBentrySTDinterwordspacing

\bibitem{PhysRevE.86.031122}
\BIBentryALTinterwordspacing
M.~L.~R. F\"urst, C.~B. Mendl, and H.~Spohn, ``Matrix-valued boltzmann equation
  for the hubbard chain,'' \emph{Phys. Rev. E}, vol.~86, p. 031122, Sep 2012.
  [Online]. Available:
  \url{https://link.aps.org/doi/10.1103/PhysRevE.86.031122}
\BIBentrySTDinterwordspacing

\bibitem{PhysRevE.88.012108}
\BIBentryALTinterwordspacing
------, ``Matrix-valued boltzmann equation for the nonintegrable hubbard
  chain,'' \emph{Phys. Rev. E}, vol.~88, p. 012108, Jul 2013. [Online].
  Available: \url{https://link.aps.org/doi/10.1103/PhysRevE.88.012108}
\BIBentrySTDinterwordspacing

\bibitem{stark2013kinetic}
M.~Stark and M.~Kollar, ``Kinetic description of thermalization dynamics in
  weakly interacting quantum systems,'' \emph{arXiv preprint arXiv:1308.1610},
  2013.

\bibitem{vidmar2016generalized}
L.~Vidmar and M.~Rigol, ``Generalized gibbs ensemble in integrable lattice
  models,'' \emph{Journal of Statistical Mechanics: Theory and Experiment},
  vol. 2016, no.~6, p. 064007, 2016.

\bibitem{d2016quantum}
L.~D'Alessio, Y.~Kafri, A.~Polkovnikov, and M.~Rigol, ``From quantum chaos and
  eigenstate thermalization to statistical mechanics and thermodynamics,''
  \emph{Advances in Physics}, vol.~65, no.~3, pp. 239--362, 2016.

\bibitem{PhysRevLett.125.180605}
\BIBentryALTinterwordspacing
M.~\ifmmode \check{Z}\else \v{Z}\fi{}nidari\ifmmode~\check{c}\else \v{c}\fi{},
  ``Weak integrability breaking: Chaos with integrability signature in coherent
  diffusion,'' \emph{Phys. Rev. Lett.}, vol. 125, p. 180605, Oct 2020.
  [Online]. Available:
  \url{https://link.aps.org/doi/10.1103/PhysRevLett.125.180605}
\BIBentrySTDinterwordspacing

\bibitem{PhysRevB.102.184304}
\BIBentryALTinterwordspacing
J.~a.~S. Ferreira and M.~Filippone, ``Ballistic-to-diffusive transition in spin
  chains with broken integrability,'' \emph{Phys. Rev. B}, vol. 102, p. 184304,
  Nov 2020. [Online]. Available:
  \url{https://link.aps.org/doi/10.1103/PhysRevB.102.184304}
\BIBentrySTDinterwordspacing

\bibitem{znidaric2021less}
M.~Znidaric, ``Less is more: more scattering leading to less resistance,''
  \emph{arXiv preprint arXiv:2109.08390}, 2021.

\bibitem{bulchandani2021onset}
V.~B. Bulchandani, D.~A. Huse, and S.~Gopalakrishnan, ``Onset of many-body
  quantum chaos due to breaking integrability,'' \emph{arXiv preprint
  arXiv:2112.14762}, 2021.

\bibitem{bertini2021finite}
B.~Bertini, F.~Heidrich-Meisner, C.~Karrasch, T.~Prosen, R.~Steinigeweg, and
  M.~{\v{Z}}nidari{\v{c}}, ``Finite-temperature transport in one-dimensional
  quantum lattice models,'' \emph{Reviews of Modern Physics}, vol.~93, no.~2,
  p. 025003, 2021.

\bibitem{PhysRevX.6.041065}
\BIBentryALTinterwordspacing
O.~A. Castro-Alvaredo, B.~Doyon, and T.~Yoshimura, ``Emergent hydrodynamics in
  integrable quantum systems out of equilibrium,'' \emph{Phys. Rev. X}, vol.~6,
  p. 041065, Dec 2016. [Online]. Available:
  \url{https://link.aps.org/doi/10.1103/PhysRevX.6.041065}
\BIBentrySTDinterwordspacing

\bibitem{PhysRevLett.117.207201}
\BIBentryALTinterwordspacing
B.~Bertini, M.~Collura, J.~De~Nardis, and M.~Fagotti, ``Transport in
  out-of-equilibrium $xxz$ chains: Exact profiles of charges and currents,''
  \emph{Phys. Rev. Lett.}, vol. 117, p. 207201, Nov 2016. [Online]. Available:
  \url{https://link.aps.org/doi/10.1103/PhysRevLett.117.207201}
\BIBentrySTDinterwordspacing

\bibitem{doyon2020lecture}
B.~Doyon, ``Lecture notes on generalised hydrodynamics,'' \emph{SciPost Physics
  Lecture Notes}, p. 018, 2020.

\bibitem{friedman2020diffusive}
A.~J. Friedman, S.~Gopalakrishnan, and R.~Vasseur, ``Diffusive hydrodynamics
  from integrability breaking,'' \emph{Physical Review B}, vol. 101, no.~18, p.
  180302, 2020.

\bibitem{durnin2020non}
J.~Durnin, M.~Bhaseen, and B.~Doyon, ``Non-equilibrium dynamics and weakly
  broken integrability,'' \emph{arXiv preprint arXiv:2004.11030}, 2020.

\bibitem{PhysRevB.103.L060302}
\BIBentryALTinterwordspacing
J.~Lopez-Piqueres, B.~Ware, S.~Gopalakrishnan, and R.~Vasseur, ``Hydrodynamics
  of nonintegrable systems from a relaxation-time approximation,'' \emph{Phys.
  Rev. B}, vol. 103, p. L060302, Feb 2021. [Online]. Available:
  \url{https://link.aps.org/doi/10.1103/PhysRevB.103.L060302}
\BIBentrySTDinterwordspacing

\bibitem{PhysRevLett.126.090602}
\BIBentryALTinterwordspacing
F.~M\o{}ller, C.~Li, I.~Mazets, H.-P. Stimming, T.~Zhou, Z.~Zhu, X.~Chen, and
  J.~Schmiedmayer, ``Extension of the generalized hydrodynamics to the
  dimensional crossover regime,'' \emph{Phys. Rev. Lett.}, vol. 126, p. 090602,
  Mar 2021. [Online]. Available:
  \url{https://link.aps.org/doi/10.1103/PhysRevLett.126.090602}
\BIBentrySTDinterwordspacing

\bibitem{PhysRevB.102.161110}
\BIBentryALTinterwordspacing
A.~Bastianello, J.~De~Nardis, and A.~De~Luca, ``Generalized hydrodynamics with
  dephasing noise,'' \emph{Phys. Rev. B}, vol. 102, p. 161110, Oct 2020.
  [Online]. Available:
  \url{https://link.aps.org/doi/10.1103/PhysRevB.102.161110}
\BIBentrySTDinterwordspacing

\bibitem{bastianello2021hydrodynamics}
A.~Bastianello, A.~De~Luca, and R.~Vasseur, ``Hydrodynamics of weak
  integrability breaking,'' \emph{arXiv preprint arXiv:2103.11997}, 2021.

\bibitem{PhysRevX.8.021030}
\BIBentryALTinterwordspacing
Y.~Tang, W.~Kao, K.-Y. Li, S.~Seo, K.~Mallayya, M.~Rigol, S.~Gopalakrishnan,
  and B.~L. Lev, ``Thermalization near integrability in a dipolar quantum
  newton's cradle,'' \emph{Phys. Rev. X}, vol.~8, p. 021030, May 2018.
  [Online]. Available: \url{https://link.aps.org/doi/10.1103/PhysRevX.8.021030}
\BIBentrySTDinterwordspacing

\bibitem{martin1984algebraic}
O.~Martin, A.~M. Odlyzko, and S.~Wolfram, ``Algebraic properties of cellular
  automata,'' \emph{Communications in mathematical physics}, vol.~93, no.~2,
  pp. 219--258, 1984.

\bibitem{schollwock2011density}
U.~Schollw{\"o}ck, ``The density-matrix renormalization group in the age of
  matrix product states,'' \emph{Annals of physics}, vol. 326, no.~1, pp.
  96--192, 2011.

\bibitem{1993CMaPh.158..127B}
A.~{Bobenko}, M.~{Bordemann}, C.~{Gunn}, and U.~{Pinkall}, ``{On two integrable
  cellular automata},'' \emph{Communications in Mathematical Physics}, vol.
  158, no.~1, pp. 127--134, Nov. 1993.

\bibitem{yang1969thermodynamics}
C.-N. Yang and C.~P. Yang, ``Thermodynamics of a one-dimensional system of
  bosons with repulsive delta-function interaction,'' \emph{Journal of
  Mathematical Physics}, vol.~10, no.~7, pp. 1115--1122, 1969.

\bibitem{takahashi2005thermodynamics}
M.~Takahashi, \emph{Thermodynamics of one-dimensional solvable models}.\hskip
  1em plus 0.5em minus 0.4em\relax Cambridge university press, 2005.

\bibitem{de2021correlation}
J.~De~Nardis, B.~Doyon, M.~Medenjak, and M.~Panfil, ``Correlation functions and
  transport coefficients in generalised hydrodynamics,'' \emph{arXiv preprint
  arXiv:2104.04462}, 2021.

\bibitem{de2019diffusion}
J.~De~Nardis, D.~Bernard, and B.~Doyon, ``Diffusion in generalized
  hydrodynamics and quasiparticle scattering,'' \emph{SciPost Phys.}, vol.~6,
  no.~4, p. 049, 2019.

\bibitem{PhysRevB.96.081118}
\BIBentryALTinterwordspacing
E.~Ilievski and J.~De~Nardis, ``Ballistic transport in the one-dimensional
  hubbard model: The hydrodynamic approach,'' \emph{Phys. Rev. B}, vol.~96, p.
  081118, Aug 2017. [Online]. Available:
  \url{https://link.aps.org/doi/10.1103/PhysRevB.96.081118}
\BIBentrySTDinterwordspacing

\bibitem{SciPostPhys.3.6.039}
\BIBentryALTinterwordspacing
B.~Doyon and H.~Spohn, ``{Drude Weight for the Lieb-Liniger Bose Gas},''
  \emph{SciPost Phys.}, vol.~3, p. 039, 2017. [Online]. Available:
  \url{https://scipost.org/10.21468/SciPostPhys.3.6.039}
\BIBentrySTDinterwordspacing

\bibitem{PhysRevLett.121.160603}
\BIBentryALTinterwordspacing
J.~De~Nardis, D.~Bernard, and B.~Doyon, ``Hydrodynamic diffusion in integrable
  systems,'' \emph{Phys. Rev. Lett.}, vol. 121, p. 160603, Oct 2018. [Online].
  Available: \url{https://link.aps.org/doi/10.1103/PhysRevLett.121.160603}
\BIBentrySTDinterwordspacing

\bibitem{fujimoto1998exact}
S.~Fujimoto and N.~Kawakami, ``Exact drude weight for the one-dimensional
  hubbard model at finite temperatures,'' \emph{Journal of Physics A:
  Mathematical and General}, vol.~31, no.~2, p. 465, 1998.

\bibitem{zotos1999finite}
X.~Zotos, ``Finite temperature drude weight of the one-dimensional spin-1/2
  heisenberg model,'' \emph{Physical review letters}, vol.~82, no.~8, p. 1764,
  1999.

\bibitem{klumper2002thermal}
A.~Kl{\"u}mper and K.~Sakai, ``The thermal conductivity of the spin-$1/2$ xxz
  chain at arbitrary temperature,'' \emph{Journal of Physics A: Mathematical
  and General}, vol.~35, no.~9, p. 2173, 2002.

\bibitem{sakai2003non}
K.~Sakai and A.~Kl{\"u}mper, ``Non-dissipative thermal transport in the massive
  regimes of the xxz chain,'' \emph{Journal of Physics A: Mathematical and
  General}, vol.~36, no.~46, p. 11617, 2003.

\bibitem{PhysRevLett.106.217206}
\BIBentryALTinterwordspacing
T.~c.~v. Prosen, ``Open $xxz$ spin chain: Nonequilibrium steady state and a
  strict bound on ballistic transport,'' \emph{Phys. Rev. Lett.}, vol. 106, p.
  217206, May 2011. [Online]. Available:
  \url{https://link.aps.org/doi/10.1103/PhysRevLett.106.217206}
\BIBentrySTDinterwordspacing

\bibitem{PhysRevLett.111.057203}
\BIBentryALTinterwordspacing
T.~c.~v. Prosen and E.~Ilievski, ``Families of quasilocal conservation laws and
  quantum spin transport,'' \emph{Phys. Rev. Lett.}, vol. 111, p. 057203, Aug
  2013. [Online]. Available:
  \url{https://link.aps.org/doi/10.1103/PhysRevLett.111.057203}
\BIBentrySTDinterwordspacing

\bibitem{PhysRevB.97.045407}
\BIBentryALTinterwordspacing
V.~B. Bulchandani, R.~Vasseur, C.~Karrasch, and J.~E. Moore, ``Bethe-boltzmann
  hydrodynamics and spin transport in the xxz chain,'' \emph{Phys. Rev. B},
  vol.~97, p. 045407, Jan 2018. [Online]. Available:
  \url{https://link.aps.org/doi/10.1103/PhysRevB.97.045407}
\BIBentrySTDinterwordspacing

\bibitem{PhysRevLett.119.020602}
\BIBentryALTinterwordspacing
E.~Ilievski and J.~De~Nardis, ``Microscopic origin of ideal conductivity in
  integrable quantum models,'' \emph{Phys. Rev. Lett.}, vol. 119, p. 020602,
  Jul 2017. [Online]. Available:
  \url{https://link.aps.org/doi/10.1103/PhysRevLett.119.020602}
\BIBentrySTDinterwordspacing

\bibitem{bouchoule2021generalized}
I.~Bouchoule and J.~Dubail, ``Generalized hydrodynamics in the 1d bose gas:
  theory and experiments,'' \emph{arXiv preprint arXiv:2108.02509}, 2021.

\bibitem{PhysRevLett.119.220604}
\BIBentryALTinterwordspacing
V.~B. Bulchandani, R.~Vasseur, C.~Karrasch, and J.~E. Moore, ``Solvable
  hydrodynamics of quantum integrable systems,'' \emph{Phys. Rev. Lett.}, vol.
  119, p. 220604, Nov 2017. [Online]. Available:
  \url{https://link.aps.org/doi/10.1103/PhysRevLett.119.220604}
\BIBentrySTDinterwordspacing

\bibitem{PhysRevLett.123.130602}
\BIBentryALTinterwordspacing
A.~Bastianello, V.~Alba, and J.-S. Caux, ``Generalized hydrodynamics with
  space-time inhomogeneous interactions,'' \emph{Phys. Rev. Lett.}, vol. 123,
  p. 130602, Sep 2019. [Online]. Available:
  \url{https://link.aps.org/doi/10.1103/PhysRevLett.123.130602}
\BIBentrySTDinterwordspacing

\bibitem{gopalakrishnan2019kinetic}
S.~Gopalakrishnan and R.~Vasseur, ``Kinetic theory of spin diffusion and
  superdiffusion in x x z spin chains,'' \emph{Physical review letters}, vol.
  122, no.~12, p. 127202, 2019.

\bibitem{PhysRevLett.111.197203}
\BIBentryALTinterwordspacing
M.~Marcuzzi, J.~Marino, A.~Gambassi, and A.~Silva, ``Prethermalization in a
  nonintegrable quantum spin chain after a quench,'' \emph{Phys. Rev. Lett.},
  vol. 111, p. 197203, Nov 2013. [Online]. Available:
  \url{https://link.aps.org/doi/10.1103/PhysRevLett.111.197203}
\BIBentrySTDinterwordspacing

\bibitem{PhysRevB.84.054304}
\BIBentryALTinterwordspacing
M.~Kollar, F.~A. Wolf, and M.~Eckstein, ``Generalized gibbs ensemble prediction
  of prethermalization plateaus and their relation to nonthermal steady states
  in integrable systems,'' \emph{Phys. Rev. B}, vol.~84, p. 054304, Aug 2011.
  [Online]. Available:
  \url{https://link.aps.org/doi/10.1103/PhysRevB.84.054304}
\BIBentrySTDinterwordspacing

\bibitem{langen2016prethermalization}
T.~Langen, T.~Gasenzer, and J.~Schmiedmayer, ``Prethermalization and universal
  dynamics in near-integrable quantum systems,'' \emph{Journal of Statistical
  Mechanics: Theory and Experiment}, vol. 2016, no.~6, p. 064009, 2016.

\bibitem{PhysRevB.95.104304}
\BIBentryALTinterwordspacing
F.~R.~A. Biebl and S.~Kehrein, ``Thermalization rates in the one-dimensional
  hubbard model with next-to-nearest neighbor hopping,'' \emph{Phys. Rev. B},
  vol.~95, p. 104304, Mar 2017. [Online]. Available:
  \url{https://link.aps.org/doi/10.1103/PhysRevB.95.104304}
\BIBentrySTDinterwordspacing

\bibitem{mallayya2019prethermalization}
K.~Mallayya, M.~Rigol, and W.~De~Roeck, ``Prethermalization and thermalization
  in isolated quantum systems,'' \emph{Physical Review X}, vol.~9, no.~2, p.
  021027, 2019.

\bibitem{lifschitz1983physical}
E.~Lifschitz and L.~Pitajewski, ``Physical kinetics,'' in \emph{Textbook of
  theoretical physics. 10}, 1983.

\bibitem{nielsen2002quantum}
M.~A. Nielsen and I.~Chuang, ``Quantum computation and quantum information,''
  2002.

\bibitem{orus2014practical}
R.~Or{\'u}s, ``A practical introduction to tensor networks: Matrix product
  states and projected entangled pair states,'' \emph{Annals of Physics}, vol.
  349, pp. 117--158, 2014.

\bibitem{fishman2020itensor}
M.~Fishman, S.~R. White, and E.~M. Stoudenmire, ``The itensor software library
  for tensor network calculations,'' 2020.

\end{thebibliography}
\end{document}